\documentclass[preprint,sort&compress,12pt]{elsarticle}

\usepackage{color}
\usepackage{axodraw}
\usepackage{amsmath,amsfonts,amssymb}

\allowdisplaybreaks

\definecolor{mygreen}{rgb}{0,1,0}
\definecolor{mygreen}{rgb}{0,.75,0}
\renewcommand{\Green}{\textcolor{mygreen}}
\definecolor{mycyan}{cmyk}{1,0,0,0}
\definecolor{mycyan}{cmyk}{.8,.15,0,0}
\definecolor{mycyan}{cmyk}{.8,.55,0,0}

\definecolor{mymagenta}{cmyk}{0,1,0,0}
\definecolor{mymagenta}{cmyk}{.15,1,0,0}

\usepackage{amssymb}

\newcounter{bla}

\journal{Computer Physics Communications}

\usepackage{hyperref}
\usepackage{supertabular}

\def\beq{\begin{equation}}
\def\eeq{\end{equation}}
\def\beqar{\begin{eqnarray}}
\def\eeqar{\end{eqnarray}}
\def\barr#1{\begin{array}{#1}}
\def\earr{\end{array}}
\def\bfi{\begin{figure}}
\def\efi{\end{figure}}
\def\btab{\begin{table}}
\def\etab{\end{table}}
\def\bce{\begin{center}}
\def\ece{\end{center}}
\def\nn{\nonumber}
\def\disp{\displaystyle}
\def\text{\textstyle}

\def\arraystretch{1.2}

\def\al{\alpha}

\def\Ga{\Gamma}

\def\eps{\epsilon}

\def\si{\sigma}

\def\refeq#1{\mbox{(\ref{#1})}}

\def\reffi#1{\mbox{Figure~\ref{#1}}}

\def\refta#1{\mbox{Table~\ref{#1}}}

\def\refse#1{\mbox{Section~\ref{#1}}}

\def\citere#1{\mbox{Ref.~\cite{#1}}}
\def\citeres#1{\mbox{Refs.~\cite{#1}}}

\newcommand{\ri}{{\mathrm{i}}}
\newcommand{\rd}{{\mathrm{d}}}

\newcommand{\ord}{\mathswitch{{\cal{O}}}}

\def\mathswitchr#1{\relax\ifmmode{\mathrm{#1}}\else$\mathrm{#1}$\fi}

\newcommand{\PW}{\mathswitchr W}

\newcommand{\PZ}{\mathswitchr Z}

\newcommand{\PH}{\mathswitchr H}
\newcommand{\Pp}{\mathswitchr p}

\newcommand{\Pb}{\mathswitchr b}

\newcommand{\Pt}{\mathswitchr t}

\newcommand{\Pep}{\mathswitchr {e^+}}
\newcommand{\Pem}{\mathswitchr {e^-}}

\newcommand{\Pj}{\mathswitchr j}
\newcommand{\Pl}{\mathswitchr \ell}

\def\mathswitch#1{\relax\ifmmode#1\else$#1$\fi}

\def\ie{i.e.\ }
\def\eg{e.g.\ }

\newcommand{\UV}{{\mathrm{UV}}}
\newcommand{\IR}{{\mathrm{IR}}}

\newcommand{\fin}{{\mathrm{fin}}}

\newcommand{\collier}{{\sc Collier}}
\newcommand{\coli}{{\tt COLI}}
\newcommand{\DD}{{\tt DD}}
\newcommand{\tensors}{{\tt tensors}}
\newcommand{\cmake}{{\sc CMake}}

\newcommand{\ina}{i_1}
\newcommand{\inb}{i_2}
\newcommand{\inc}{i_3}
\newcommand{\ind}{i_4}
\newcommand{\ine}{i_5}
\newcommand{\ing}{i_6}

\newcommand{\np}{n_p}

\newcommand{\sst}{\scriptstyle}
\def\nl{\nonumber\\}
\def\nls{\nonumber\\[1ex]}
\def\nlss{\nonumber\\*[1ex]}

\newcommand{\nTn}{{\tt Nn}}

\newlength{\parwidth}\newlength{\colonewidth}%
\newlength{\restpageheight}

\makeatletter
\newcommand{\cpcsuptable}[2]
{\settowidth{\colonewidth}{#1}\setlength{\parwidth}{\textwidth}%
\addtolength{\parwidth}{-\colonewidth}\addtolength{\parwidth}{-3em}%
\bce
\setlength{\restpageheight}{\@colroom}\addtolength{\restpageheight}{-\pagetotal}
\ifdim \restpageheight<20pt \pagebreak\fi
\begin{supertabular}[l]{p{\colonewidth}@{ }c@{ }p{\parwidth}}
#2
\end{supertabular}%
\ece
}%
\makeatother

\makeatletter
\renewcommand{\cpcsuptable}[2]%
{\settowidth{\colonewidth}{#1}\setlength{\parwidth}{\textwidth}%
\addtolength{\parwidth}{-\colonewidth}\addtolength{\parwidth}{-3em}%
\begin{center}
\setlength{\restpageheight}{\@colroom}\addtolength{\restpageheight}{-\pagetotal}%
\ifdim \restpageheight<20pt \pagebreak\fi
\begin{supertabular}[l]{p{\colonewidth}@{ : }p{\parwidth}}%
#2
\end{supertabular}%
\end{center}
}%
\makeatother

\newcommand{\cpcsub}[1]
{%
\setlength{\parwidth}{\textwidth}\addtolength{\parwidth}{-2.1em}%
\bce
\begin{tabular}[t]{@{}p{\parwidth}@{}}
#1
\end{tabular}
\ece
}%

   {\end{description}}

\def\draftdate{\relax}
\def\mda{\relax}
\def\mua{\relax}
\def\mla{\relax}
\def\draft{
\def\thtystars{******************************}
\def\sixtystars{\thtystars\thtystars}
\typeout{}
\typeout{\sixtystars**}
\typeout{* Draft mode!
         For final version remove \protect\draft\space in source file *}
\typeout{\sixtystars**}
\typeout{}
\def\draftdate{\today}
\def\mua{\marginpar[\boldmath\hfil$\uparrow$]%
                   {\boldmath$\uparrow$\hfil}%
                    \typeout{marginpar: $\uparrow$}\ignorespaces}
\def\mda{\marginpar[\boldmath\hfil$\downarrow$]%
                   {\boldmath$\downarrow$\hfil}%
                    \typeout{marginpar: $\downarrow$}\ignorespaces}
\def\mla{\marginpar[\boldmath\hfil$\rightarrow$]%
                   {\boldmath$\leftarrow $\hfil}%
                    \typeout{marginpar: $\leftrightarrow$}\ignorespaces}
\def\Mua{\marginpar[\boldmath\hfil$\Uparrow$]%
                   {\boldmath$\Uparrow$\hfil}%
                    \typeout{marginpar: $\uparrow$}\ignorespaces}
\def\Mda{\marginpar[\boldmath\hfil$\Downarrow$]%
                   {\boldmath$\Downarrow$\hfil}%
                    \typeout{marginpar: $\downarrow$}\ignorespaces}
\def\Mla{\marginpar[\boldmath\hfil$\Rightarrow$]%
                   {\boldmath$\Leftarrow $\hfil}%
                    \typeout{marginpar: $\leftrightarrow$}\ignorespaces}
\overfullrule 5pt
\oddsidemargin -15mm
\marginparwidth 29mm
}

\begin{document}

\begin{frontmatter}



\title{\collier: {a fortran-based  Complex One-Loop LIbrary in
  Extended Regularizations}}
\tnotetext[mytitlenote]{The program is available from  
  \href{http://collier.hepforge.org/}{\mbox{http://collier.hepforge.org}}.}


\author[ad]{Ansgar Denner}
\ead{ansgar.denner@physik.uni-wuerzburg.de}

\author[sd]{Stefan Dittmaier}
\ead{stefan.dittmaier@physik.uni-freiburg.de}

\author[lh]{Lars Hofer}
\ead{hofer@ecm.ub.edu}


\address[ad]{Universit\"at W\"urzburg, 
Institut f\"ur Theoretische Physik und Astrophysik, \\
D-97074 W\"urzburg, Germany}
\address[sd]{Albert-Ludwigs-Universit\"at Freiburg, Physikalisches Institut, 
\\ D-79104 Freiburg, Germany}
\address[lh]{Department de F\'isica Qu\`antica i Astrof\'isica (FQA), \\
Institut de Ci\`encies del Cosmos (ICCUB), \\
Universitat de Barcelona (UB),
Mart\'i Franqu\`es 1, \\
E-08028 Barcelona, Spain}

\begin{abstract}
  We present the library \collier\ for the numerical evaluation of
  one-loop scalar and tensor integrals in perturbative relativistic
  quantum field theories. The code provides numerical results for
  arbitrary tensor and scalar integrals for scattering processes in
  general quantum field theories. For tensor integrals either the
  coefficients in a covariant decomposition or the tensor components
  themselves are provided. \collier\ supports complex masses, which
  are needed in calculations involving unstable particles.
  Ultraviolet and infrared singularities are treated in dimensional
  regularization. For soft and collinear singularities 
  mass regularization is available as an alternative.
\end{abstract}

\begin{keyword}
NLO computations; radiative corrections; one-loop integrals; higher orders; tensor reduction;
scalar integrals

\end{keyword}

\end{frontmatter}



\noindent
{\bf PROGRAM SUMMARY}

\begin{small}
\noindent
{\em Manuscript Title:}  \collier: {a fortran-based Complex One-Loop LIbrary in
  Extended Regularizations} \\
{\em Authors:} Ansgar Denner, Stefan Dittmaier, Lars Hofer                                                \\
{\em Program Title: }\collier                                          \\
{\em Journal Reference:}                                      \\
{\em Catalogue identifier:}                                   \\
{\em Licensing provisions:}         GNU GPL version 3
\\
{\em Programming language:} Fortran95                                  \\
{\em Computer:} any with a Fortran95 compiler                             \\
{\em Operating system:}      Linux, Mac~OS~X  
                             \\
{\em RAM:} required RAM is insignificant                             \\
{\em Number of processors used:}     one                         \\
{\em Supplementary material:}   none 
                              \\
{\em Keywords:} NLO computations, radiative corrections, one-loop
integrals, tensor reduction, scalar integrals
\\
{\em Classification:}\\                 
 4.4 Feynman diagrams,  11.1 General, High Energy Physics and Computing\\
{\em External routines/libraries:}        none
                     \\
{\em Subprograms used:}                 none
                       \\
{\em Nature of problem:} \\
Evaluation of general one-loop multi-leg
scalar and tensor integrals occurring in the calculation of one-loop
corrections to scattering amplitudes in relativistic quantum field theories
   \\
{\em Solution method:}\\
Scalar integrals are evaluated using explicit analytical expressions.
Tensor integrals are numerically reduced to scalar integrals via
different methods. Depending on the specific kinematical variables,
an appropriate method is automatically 
chosen to optimize the resulting numerical accuracy.
   \\
{\em Restrictions:}\\
real momenta
  \\
{\em Additional comments:}       none
 \\
{\em Running time:}\\
Depends on the nature of the problem.
Typically the CPU time needed 
is comparable or smaller than the time required
for evaluating the remaining building blocks of a one-loop amplitude.
   \\

\end{small}

\clearpage

\section{Introduction}
\label{se:intro}

The exploitation of experiments at high-energy colliders like the LHC
heavily relies on theoretical predictions for scattering processes
within quantum field theories, which are basically evaluated within
perturbation theory.  Next-to-leading-order (NLO) perturbative
corrections of the strong interaction are a crucial ingredient for
decent predictions, but also NLO electroweak corrections are required
in many cases
(see, for instance, \citeres{Campbell:2013qaa,Butterworth:2014efa}).

\begin{sloppypar}
In the calculation of NLO QCD corrections huge progress has been made
in recent years (see \eg 
\citeres{Campbell:2013qaa,Butterworth:2014efa,Bern:2008ef,Binoth:2010ra,AlcarazMaestre:2012vp}), and
automated tools have
become available such as 
{\sc Blackhat}~\cite{Berger:2008sj}, 
{\sc NGluon}~\cite{Badger:2010nx}, 
{\sc HELAC-NLO}~\cite{Bevilacqua:2011xh},
{\sc GoSam}~\cite{Cullen:2011ac}, and
{\sc MadGraph5\_aMC@NLO}~\cite{Alwall:2014hca}.
For the calculation of the one-loop
matrix elements different approaches have been developed and
implemented like 
{\sc CutTools}~\cite{Ossola:2007ax}, 
{\sc HELAC-1LOOP}~\cite{vanHameren:2009dr}, 
{\sc Samurai}~\cite{Mastrolia:2010nb}, 
{\sc Madloop}~\cite{Hirschi:2011pa}, 
{\sc OpenLoops}~\cite{Cascioli:2011va}, 
or {\sc Recola}~\cite{Actis:2012qn}.
Electroweak NLO corrections, 
whose automation for multi-leg computations started with the development of the recursive
generator 
{\sc Recola}~\cite{Actis:2012qn}, have recently been implemented also in
{\sc MadGraph5\_aMC@NLO}~\cite{Frixione:2014qaa,Frixione:2015zaa} and
{\sc OpenLoops}~\cite{Kallweit:2014xda,Kallweit:2015dum}. 
Finally, the packages {\sc
  FeynArts}~\cite{Hahn:2000kx},
{\sc  FeynCalc}~\cite{Mertig:1990an,Shtabovenko:2016sxi}, and 
{\sc FormCalc}~\cite{Hahn:1998yk,Agrawal:2012cv,Nejad:2013ina}
provide flexible tools to evaluate NLO amplitudes within and
beyond the Standard Model, with more emphasis on generality than on high multiplicities.
\end{sloppypar}

The great progress in the evaluation of multi-leg one-loop amplitudes
was triggered both by systematic improvements in the traditional
Feynman-diagrammatic approach and by the development of new
field-theoretical techniques based on generalized unitarity relations.
In the latter approaches \cite{Bern:1994zx, Bern:1994cg,
  Britto:2004nc, Ossola:2006us, Ellis:2007br,Giele:2008ve,
  Ellis:2008ir}, 
one-loop amplitudes are directly expressed in terms
of scalar integrals. 
The direct reduction of the amplitude to
the fixed set of scalar integrals leads to numerical problems in
specific regions of phase space, which are usually overcome by
resorting to quadruple precision in the numerical calculation. The
Feynman-diagrammatic approach and also the recent recursive methods
\cite{Cascioli:2011va,Actis:2012qn,vanHameren:2009vq} instead rely on
tensor integrals. This allows to adapt the reduction method to the
respective region of phase space, and an optimal choice avoids
numerical instabilities to a large extent.  The \collier\ library
presented here represents a comprehensive tool for the evaluation of
scalar and tensor integrals that is applicable in either of the two
complementary types of approaches.

The reduction of tensor integrals to a small set of basic integrals
goes back to Brown and Feynman \cite{Brown:1952eu}, Melrose
\cite{Melrose:1965kb}, and Passarino and
Veltman~\cite{Passarino:1978jh}. 
Over the decades, the methods have been refined and
improved by various authors
\cite{vanNeerven:1983vr,vanOldenborgh:1989wn,
  Bern:1992em,Denner:1991kt,
Binoth:1999sp,Denner:2002ii,Beenakker:2002nc, 
Dittmaier:2003bc,
Duplancic:2003tv,
Giele:2004iy,Giele:2004ub,%
Binoth:2005ff,
Denner:2005nn,
Fleischer:2010sq}. %
The complete set of reduction methods presented in
\citeres{Denner:2002ii,Denner:2005nn} serves as the basis of the code
\collier.

Eventually, the reduction of tensor integrals leads to a set of scalar
integrals, which were systematically investigated for the first time
in \citere{'tHooft:1978xw} by 't Hooft and Veltman.
Subsequently, results were published for
specific IR-singular cases both in mass and dimensional regularization
\cite{Bern:1992em,Beenakker:1988jr,Bern:1993kr}, and a simplified
result for the general 4-point function with real masses was derived
\cite{Denner:1991qq}. For processes with unstable particles, in
particular in the complex-mass scheme
\cite{Denner:1999gp,Denner:2005fg}, expressions valid for complex masses of internal
particles are required. A corresponding code for the regular
4-point function has
been published in \citere{Nhung:2009pm}.  More compact results for
the general scalar 4-point function with complex masses were presented in
\citere{Denner:2010tr}, where results for all relevant 
(regular and IR-singular) cases in
scattering processes, both in mass and dimensional regularization, have been presented. These results have been
encoded in \collier, while the expressions implemented for the scalar 2- and
3-point functions are based on
\citeres{Dittmaier:2003bc,'tHooft:1978xw}.

Several integral libraries are already available 
for the calculation of one-loop scalar and tensor integrals: {\sc FF} \cite{vanOldenborgh:1990yc}, {\sc
  LoopTools} \cite{Hahn:1998yk}, {\sc QCDLoop} \cite{Ellis:2007qk},
  {\sc OneLOop} \cite{vanHameren:2010cp}, {\sc Golem95C}
\cite{Cullen:2011kv},
{\sc PJFry} \cite{Fleischer:2010sq}, and {\sc Package-X} \cite{Patel:2015tea}. The \collier\ library presented here 
  includes the complete set of tensor integrals necessary for
  processes with complex masses 
with no a-priori restriction on the number of external particles.

\collier\
has been developed and applied in the course of several cutting edge
NLO QCD + electroweak calculations to scattering processes, comprising
for instance
$\Pep\Pem\to\PW\PW\to4\,$fermions~\cite{Denner:2005fg,Denner:2005es},
$\PH\to\PW\PW/\PZ\PZ\to4\,$fermions~\cite{Bredenstein:2006rh,Bredenstein:2006ha},
$\Pp\Pp\to\Pt\bar\Pt\Pb\bar\Pb$~\cite{Bredenstein:2008zb,Bredenstein:2009aj,Bredenstein:2010rs,Cascioli:2013era},
$\Pp\Pp\to\PW\PW\Pb\bar\Pb$~\cite{Denner:2010jp,Denner:2012yc,Cascioli:2013wga},
$\Pp\Pp\to2\Pl{+}{\le}2\Pj$ \cite{Denner:2014ina,Kallweit:2015dum},
$\Pp\Pp\to\PW\PW\to4\ell$ in double-pole approximation~\cite{Billoni:2013aba},
$\Pp\Pp\to\PW{+}{\le}3\Pj$ \cite{Kallweit:2014xda},
$\Pp\Pp\to\Pt\bar\Pt\Pj\Pj$~\cite{Hoeche:2014qda},
$\Pp\Pp\to\PW\PW\Pb\bar\Pb\PH$~\cite{Denner:2015yca},
$\Pp\Pp\to\mu^+\mu^-\Pep\Pem$~\cite{Biedermann:2016yvs},
and various applications with lower multiplicities in the final
state. 
The code has also been used to calculate one-loop amplitudes
in NNLO calculations \cite{Grazzini:2013bna,Cascioli:2014yka,Gehrmann:2014fva},
where the evaluation of one-loop integrals for collinear and/or soft external momenta is required.
It is further 
integrated in the NLO generators
{\sc OpenLoops}~\cite{Cascioli:2011va} and {\sc
  Recola}~\cite{Actis:2012qn}. 

This article is organized as follows: 
After setting the relevant conventions 
used by \collier\ 
in \refse{se:conventions},
we outline the methodology applied to calculate the tensor integrals
in \refse{se:methods}.
Section~\ref{se:structure} describes the internal structure of
the \collier\  library and Section~\ref{subsec:Usage} its actual usage.
Our conclusions are given in \refse{se:concl}.
Finally, \ref{App:MomInv} provides some further details 
on the kinematical input used to define one-loop integrals.

\section{Conventions}
\label{se:conventions}

\begin{sloppypar}
We consistently 
use the conventions of \citeres{Denner:2005nn,Denner:2010tr}.  
The methods used for the reduction of tensor
integrals have been described in \citeres{Denner:2002ii,Denner:2005nn}, while
the implemented results for the scalar 4-point functions can be found in
\citere{Denner:2010tr}, those for the scalar 1-, 2-, and 3-point
functions are based on \citeres{Dittmaier:2003bc,'tHooft:1978xw}.
\end{sloppypar}

In $D$ dimensions, 
one-loop tensor $N$-point integrals have the general form
\beq
\label{tensorint}
T^{N,\mu_{1}\ldots\mu_{P}}(p_{1},\ldots,p_{N-1},m_{0},\ldots,m_{N-1})=
\displaystyle{\frac{(2\pi\mu)^{4-D}}{\ri\pi^{2}}\int \rd^{D}q\,
\frac{q^{\mu_{1}}\cdots q^{\mu_{P}}}
{N_0N_1\ldots N_{N-1}}}
\eeq
with the denominator factors
\beq \label{D0Di}
N_{k}= (q+p_{k})^{2}-m_{k}^{2}+\ri\epsilon, \qquad k=0,\ldots,N-1 ,
\qquad p_0=0,
\eeq
where $\ri\epsilon$ $(\eps>0)$ is an infinitesimally small imaginary
part.  For $P=0$, \ie with a factor~1 instead of
integration momenta in the numerator of the
loop integral, \refeq{tensorint} defines the scalar $N$-point integral
$T^N_0$.  Following the notation of \citere{'tHooft:1978xw}, we set
$T^{1}= A,$ $T^{2}= B,$ $T^{3}= C,$ $T^{4}= D$, $T^{5}= E,$ $T^{6}=
F,$ and $T^{7}=G$.  

In order to be able to write down tensor decompositions in a
concise way, we use a notation for the basic tensor structures 
in which curly brackets denote symmetrization
with respect to Lorentz indices in such a way that all non-equivalent
permutations of the Lorentz indices on metric tensors $g$ and a
generic momentum $p$ contribute with weight one. In covariants
with $\np$ momenta $p_{i_j}^{\mu_j}$ $(j=1,\dots,\np$) only one
representative out of the $\np!$ permutations of the indices $i_j$ is
kept.  Thus, we have for example
\beqar
\{p\ldots p\}^{\mu_1\ldots\mu_P}_{i_1\ldots i_P} &=& 
p_{i_1}^{\mu_1}\ldots p_{i_P}^{\mu_P},
\nn\\[.3em]
\{g p\}_{i_1}^{\mu\nu\rho} &=& 
g^{\mu\nu}p_{i_1}^{\rho}+g^{\nu\rho}p_{i_1}^{\mu}+g^{\mu\rho}p_{i_1}^{\nu},
\nn\\[.3em]
\{g pp \}^{\mu\nu\rho\si}_{i_1 i_2} &=& 
 g^{\mu\nu}p_{i_1}^{\rho}p_{i_2}^{\si}
+g^{\mu\rho}p_{i_1}^{\si}p_{i_2}^{\nu}
+g^{\mu\si}p_{i_1}^{\nu}p_{i_2}^{\rho}\nl&&{}
+g^{\nu\rho}p_{i_1}^{\si}p_{i_2}^{\mu}
+g^{\rho\si}p_{i_1}^{\nu}p_{i_2}^{\mu}
+g^{\nu\si}p_{i_1}^{\rho}p_{i_2}^{\mu},
\nn\\[.3em]
\{gg\}^{\mu\nu\rho\si} &=& g^{\mu\nu}g^{\rho\si}+g^{\mu\si}g^{\nu\rho}
+g^{\mu\rho}g^{\nu\si}.
\label{eq:covbrace}
\eeqar
These basic tensor structures can be recursively defined according to
\beqar
\{p\ldots p\}^{\mu_1\ldots\mu_P}_{i_1\ldots i_P} &=& 
p_{i_1}^{\mu_1}\ldots p_{i_P}^{\mu_P},\label{eq:MomTen}
\\[.3em]
\{\underbrace{g \ldots g}_n  p\ldots p\}^{\mu_1\ldots\mu_P}_{i_{2n+1}\ldots i_P}
&=& \frac{1}{n}
\sum_{\substack{k,l=1\\k<l}}^P \, g^{\mu_k\mu_l}
\{\underbrace{g \ldots g}_{n-1}  p\ldots p\}^{\mu_1\ldots\mu_{k-1}\mu_{k+1}\ldots\mu_{l-1}\mu_{l+1}\ldots\mu_P}_{i_{2n+1}\ldots i_P}.
\nn
\eeqar

We decompose the general tensor integral into Lorentz-covariant
structures as
\beqar
T^{N,\mu_1\ldots\mu_P} &=& 
\sum_{n=0}^{\left[\frac{P}{2}\right]} \,
\sum_{i_{2n+1},\ldots,i_P=1}^{N-1} \,
\{\underbrace{g \ldots g}_n  p\ldots p\}^{\mu_1\ldots\mu_P}_{i_{2n+1}\ldots i_P}
\, T^N_{\underbrace{\sst 0\ldots0}_{2n} i_{2n+1}\ldots i_{P}}
\nn\\
&=&\sum_{i_1,\ldots,i_P=1}^{N-1} 
p^{\mu_1}_{i_1}\ldots p^{\mu_P}_{i_P} T^N_{i_1\ldots i_P}
+ \sum_{i_3,\ldots,i_{P}=1}^{N-1} 
\{g p\ldots p\}^{\mu_1\ldots\mu_P}_{i_3\ldots i_P} T^N_{00i_3\ldots i_{P}}
\nl&&{}
+ \sum_{i_5,\ldots,i_{P}=1}^{N-1} 
\{g g p\ldots p\}^{\mu_1\ldots\mu_P}_{i_5 \ldots i_P} T^N_{0000i_5\ldots i_{P}}
+\ldots
\nl&&{}+
\left\{
\barr{cc}\disp
\sum_{i_{P}=1}^{N-1} 
\{g \ldots g p\}_{i_P}^{\mu_1\ldots\mu_P} 
T^N_{\underbrace{\sst0\ldots 0}_{P-1} i_{P}},
& \mbox{ for } P\ \mathrm{odd,} \\
\{g \ldots g \}^{\mu_1\ldots\mu_P} 
T^N_{\underbrace{\sst0\ldots 0}_{P}},
& \mbox{ for } P\ \mathrm{even,} 
\earr\right.\label{eq:decomp}
\eeqar
where $\left[{P}/{2}\right]$ is the largest integer number smaller 
or equal to ${P}/{2}$.  For each metric tensor in the Lorentz-covariant 
tensor structure, the corresponding coefficient carries an index pair ``00''
and for each momentum $p_{i_r}$ it carries the corresponding index
$i_r$. By definition, 
the tensor coefficients $T^N_{i_1\ldots i_P}$ are 
totally symmetric in the indices $i_1,\ldots,i_P$.

For tensor integrals up to rank six, the decompositions more explicitly read
\begin{align}
T^{N,\mu}=&\sum_{\ina=1}^{N-1} p_{\ina}^{\mu}T^N_{\ina}, \qquad
T^{N,\mu\nu}=\sum_{\ina,\inb=1}^{N-1} p_{\ina}^{\mu}p_{\inb}^{\nu}T^N_{\ina\inb}
+g^{\mu\nu}T^N_{00},\nls
T^{N,\mu\nu\rho}=&\sum_{\ina,\inb,\inc=1}^{N-1} p_{\ina}^{\mu}p_{\inb}^{\nu}p_{\inc}^{\rho}T^N_{\ina\inb\inc}
+\sum_{\ina=1}^{N-1}\{g p\}_{\ina}^{\mu\nu\rho} T^N_{00\ina}, 
\nls
T^{N,\mu\nu\rho\si} =& 
\sum_{\ina,\inb,\inc,\ind=1}^{N-1} p_{\ina}^{\mu}p_{\inb}^{\nu}p_{\inc}^{\rho}p_{\ind}^\si T^N_{\ina\inb\inc\ind}
+\sum_{\ina,\inb=1}^{N-1}
\{g pp\}_{\ina\inb}^{\mu\nu\rho\si}T^N_{00\ina\inb}
\nls&
+\{g g\}^{\mu\nu\rho\si} T^N_{0000},\nl
T^{N,\mu\nu\rho\si\tau} = &
\sum_{\ina,\inb,\inc,\ind,\ine=1}^{N-1}
p_{\ina}^{\mu}p_{\inb}^{\nu}p_{\inc}^{\rho}p_{\ind}^\si p_{\ine}^\tau T^N_{\ina
\inb\inc\ind\ine}
\nlss&{}
+\sum_{\ina,\inb,\inc=1}^{N-1}
\{g ppp\}_{\ina\inb\inc}^{\mu\nu\rho\si\tau}
T^N_{00\ina\inb\inc}
+\sum_{\ina=1}^{N-1} 
\{g g p\}_{\ina}^{\mu\nu\rho\si\tau} 
T^N_{0000\ina},\nls
T^{N,\mu\nu\rho\si\tau\al} =& 
\sum_{\ina,\inb,\inc,\ind,\ine,\ing=1}^{N-1}
p_{\ina}^{\mu}p_{\inb}^{\nu}p_{\inc}^{\rho}p_{\ind}^\si 
p_{\ine}^\tau p_{\ing}^\al T^N_{\ina\inb\inc\ind\ine\ing}
\nlss&{}
+\sum_{\ina,\inb,\inc,\ind=1}^{N-1}
\{g pppp\}_{\ina\inb\inc\ind}^{\mu\nu\rho\si\tau\al}
T^N_{00\ina\inb\inc\ind}
\nls&{}
+\sum_{\ina,\inb=1}^{N-1} 
\{g g pp\}_{\ina\inb}^{\mu\nu\rho\si\tau\al} 
T^N_{0000\ina\inb}
+ 
\{g g g\}^{\mu\nu\rho\si\tau\al} 
T^N_{000000}.
\end{align}

UV- or IR-singular integrals are represented in dimensional
regularization, where $D=4-2\eps$, as
\beqar
\label{eq:normalization}
T^N&=&
\tilde{T}^N_{\fin}
+  a^{\UV} \left(\Delta_{\UV}+\ln\frac{\mu^2_\UV}{Q^2}\right)
\nl&&{}
+ a^{\IR}_2 \left(\Delta^{(2)}_{\IR}+\Delta^{(1)}_{\IR}\ln\frac{\mu^2_\IR}{Q^2}+\frac{1}{2}\ln^2\frac{\mu^2_\IR}{Q^2}\right)
+ \tilde{a}^{\IR}_1 \left(\Delta^{(1)}_{\IR}+\ln\frac{\mu^2_\IR}{Q^2}\right)
\hspace{2em}\nl
&=&
T^N_{\fin}(\mu^2_\UV,\mu^2_\IR)
+  a^{\UV} \Delta_{\UV}
+ a^{\IR}_2 \left(\Delta^{(2)}_{\IR}+\Delta^{(1)}_{\IR}\ln\mu^2_\IR\right)
+ a^{\IR}_1 \Delta^{(1)}_{\IR}
\hspace{2em}
\eeqar
with
\beqar
\Delta_{\UV} &=& \frac{c(\eps_\UV)}{\eps_{\UV}}, \quad c(\eps) = \Gamma(1+\eps)(4\pi)^\eps,
\label{eq:normalizationUV}
\nonumber\\
\Delta^{(2)}_{\IR} &=& \frac{c(\eps_\IR)}{\eps_{\IR}^2}, \quad
               \Delta^{(1)}_{\IR} = \frac{c(\eps_\IR)}{\eps_{\IR}}.
\label{eq:normalizationIR}
\eeqar
We make
explicit all UV and IR poles as well as the terms involving the
corresponding mass scales $\mu_\UV$ and $\mu_\IR$.
We further factor out the term
$c(\eps)=\Gamma(1+\eps)(4\pi)^\eps=1+\mathcal{O}(\eps)$ and absorb it
into the definitions of $\Delta_{\UV}$, $\Delta^{(2)}_{\IR}$, and
$\Delta^{(1)}_{\IR}$. 
In order to avoid logarithms of dimensionful quantities in the first equation in \refeq{eq:normalization}, 
we have singled out an auxiliary scale $Q$ which is implicitly fixed by the masses and momenta entering the respective loop integral.  
The output delivered by \collier\ 
corresponds to the last line of \refeq{eq:normalization}, including
the terms proportional to $a^{\UV}$, $a^{\IR}_2$, and $a_1^{\IR}$. The
parameters $\mu^2_\UV$, $\mu^2_\IR$,
$\Delta_{\UV}$, $\Delta^{(2)}_{\IR}$, and $\Delta^{(1)}_{\IR}$ can be
chosen freely by the user, 
but do not influence UV- and IR-finite quantities.
Note that we distinguish between singularities of IR
and UV origin. By default $\Delta_{\UV}$,
$\Delta^{(2)}_{\IR}$, and $\Delta^{(1)}_{\IR}$ are set to zero and
the output equals $T^N_{\fin}(\mu^2_\UV,\mu^2_\IR)$. Setting the
$\Delta$'s different from zero, the impact of the poles 
in $\epsilon$ can be numerically simulated in the results.

By default IR- and UV-singular integrals are calculated in dimensional
regularization. Collinear singularities can also be regularized with
masses. To this end, the corresponding masses, called $\overline{m}_i$
in the following, must be declared {\it small\/} in the initialization.
Moreover, in all subroutine calls the respective mass parameters must have
exactly the same (not necessarily small) numerical value as specified
in the initialization.
The {\it small\/}
masses are treated as infinitesimally small in the scalar and tensor
functions, and only in mass-singular logarithms 
the finite values
are kept.  For soft singularities of Abelian type, \ie when
$a^{\IR}_2=0$ and collinear singularities are regularized with masses
$\overline{m}_i$, the parameter $\mu_\IR$ can be interpreted as an
infinitesimal photon or gluon mass after setting the parameter
$\Delta^{(1)}_{\IR}$ to zero.%
\footnote{Formally this means that the scales obey the
hierarchy $\mu_\IR\ll \overline{m}_i\ll M$ in the analytical derivation of the
integrals, where $M$ is any other scale involved in the calculation.}

Varying the parameters  $\mu^2_\UV$, $\mu^2_\IR$,
$\Delta_{\UV}$, $\Delta^{(2)}_{\IR}$, and $\Delta^{(1)}_{\IR}$
allows to check the cancellation of
singularities. Moreover, choosing appropriate values for 
$\Delta_{\UV}$, $\Delta^{(2)}_{\IR}$, and $\Delta^{(1)}_{\IR}$
allows the user to switch to different
conventions concerning the extraction of the prefactor $c(\eps)$ in
\refeq{eq:normalizationUV}. 
For instance, in \citere{Ellis:2007qk} the
$\eps$-dependent prefactor in the one-loop integral is $\pi^\eps/r_\Ga$ with
$r_\Ga=\Ga^2(1-\eps)\Ga(1+\eps)/\Ga(1-2\eps)$ in contrast to the prefactor
$(2\pi)^{2\eps}$ in our convention \refeq{tensorint}. We thus have to replace our factor $c(\eps)$ by
\beq
\frac{c(\eps)}{r_\Ga(4\pi)^\eps} =
\frac{\Gamma(1+\eps)}{r_\Ga}=\frac{\Ga(1-2\eps)}{\Ga^2(1-\eps)}=1+\eps^2\frac{\pi^2}{6}+\ord{(\eps^3)},
\eeq
in order to obtain the singular integrals in the conventions of \citere{Ellis:2007qk}.
This is equivalent to the recipe of replacing
$\Delta^{(2)}_{\IR}\to\Delta^{(2)}_{\IR}+\pi^2/6$ while keeping 
$\Delta_{\UV}$ and $\Delta^{(1)}_{\IR}$ unchanged in our calculation.
After this change, our parameters
$\Delta_{\UV}$, $\Delta^{(2)}_{\IR}$, and $\Delta^{(1)}_{\IR}$
simply correspond to the poles $1/\eps$, $1/\eps^2$, and $1/\eps$ of
\citere{Ellis:2007qk}, respectively.

\section{Implemented methods}
\label{se:methods}

\subsection{Calculation of tensor coefficients}
\label{sec:calc-tensor-coefficients}

The method used to evaluate a tensor integral depends on the number
$N$ of its propagators. For $N=1,2$, we use explicit numerically stable
expressions~\cite{Passarino:1978jh,Denner:2005nn}.

\begin{sloppypar}
For $N=3,4$, all tensor integrals are numerically 
reduced to the basic scalar
integrals, which are calculated using analytical expressions
given in \citeres{Dittmaier:2003bc,'tHooft:1978xw,Denner:2010tr}.  By
default, the reduction is performed
via standard Passarino--Veltman reduction \cite{Passarino:1978jh}. In
regions of the phase space where a Gram determinant becomes small,
Passarino--Veltman reduction becomes unstable.  In these regions we
use the dedicated recursive expansion methods described in
\citere{Denner:2005nn} plus some additional variants. All
these methods have been implemented in {\collier} to arbitrary order
in the expansion parameter. In order to decide on the method to use
for a certain phase-space point, the following procedure is applied:
\begin{enumerate}
\item
Passarino--Veltman reduction is used by default, and
its reliability is assessed by first assigning appropriate
accuracies for the scalar integrals and subsequently
estimating the error propagation during the reduction.
If the resulting error estimate $\Delta T^N(\widehat{P})$
for the integrals of the highest required rank $\widehat{P}$
is smaller than a predefined precision tag $\eta_{\textrm{req}}$ (required precision),
the result is kept and returned to the user.
\item In case step~1 does not provide sufficient accuracy, which
  typically means that the tensor integral involves small Gram
  determinants of external momenta, \collier\ switches to dedicated
  expansions.  In order to decide which expansion is appropriate, an
  a-priori error estimate $\Delta T^N_{\mathrm{prelim}}(P)$ for the
  coefficients $T^N_{i_{1}\ldots i_{P}}$ 
is constructed up to the highest required rank $\widehat{P}$ 
  for the different methods. The estimates
  are based on an assessment of the expected accuracy of the expansion
  and a simplified propagation of errors from the required scalar
  integrals.  The expansion method with the smallest $\Delta
  T^N_{\mathrm{prelim}}(\widehat{P})$ is chosen.
  During the actual calculation
  of this expansion a more realistic precision $\Delta T^N(P)$ is
  assessed by analysing the correction of the last iteration.  If the
  predefined precision tag $\eta_{\textrm{req}}$ is reached, the
  result is kept and returned to the user.  Otherwise the expansion
  stops if either a predefined iteration depth is reached, or if the
  accuracy does not increase anymore from one iteration step to the
  next.
\item
If step~2 does not provide sufficient accuracy for the selected method,
 it is repeated for the other expansion methods that promise
convergence by sufficiently small $\Delta T^N_{\mathrm{prelim}}(\widehat{P})$.
If the predefined precision tag is reached after any of these
repetitions, the result is kept and returned to the user.
\item
If neither Passarino--Veltman reduction nor any of
the tried expansions delivers results that match
the target accuracy, the results of the method with
the smallest error estimate $\Delta T^N(P)$ is
returned to the user.
\end{enumerate}
In this way stable results are obtained for almost all phase-space
points, 
ensuring reliable Monte Carlo integrations.
\end{sloppypar}

For $N=5,6$, tensor integrals are directly reduced to integrals with
lower rank and lower $N$ following
\citeres{Denner:2002ii,Denner:2005nn}, \ie 
without involving inverse Gram determinants. 
For $N\ge7$ a modification of the reduction of 6-point tensor
integrals is applied as described in Section~7 of \citeres{Denner:2005nn}
$[$see text after (7.10) there$]$.

\subsection{Calculation of full tensors}

While the methods described so far are formulated in the literature in
terms of the Lorentz-invariant coefficients $T^N_{i_1\ldots i_P}$, a
new generation of NLO generators, such as {\sc OpenLoops} and {\sc
  Recola}, needs the components of the full tensors
$T^{N,\mu_1\cdots\mu_{P}}$. To this end, an efficient algorithm has
been implemented in {\collier} to construct the tensors
$T^{N,\mu_1\cdots\mu_{P}}$ from the coefficients $T^N_{i_1\ldots
  i_P}$. It performs a recursive calculation of those tensor
structures (\ref{eq:MomTen}) which
are built exclusively from momenta.
Non-vanishing components of tensor structures involving metric
tensors are then obtained recursively by adding pairwise equal Lorentz
indices to tensor structures with less metric tensors,
taking into account the combinatorics of the indices and the signs
induced by the metric tensors.
The relevant combinatorial
factors are calculated and tabulated during the initialization of
{\collier}.
\begin{table}[t]
\let\Green\relax
\let\Red\relax
\let\Blue\relax
\small
  $$\begin{array}{c@{\quad}|c@{\quad}c@{\quad}c@{\quad}c@{\quad}c@{\quad}c@{\quad}c}
{\rm coefficients~for}
             &  \Blue{\widehat{P}=0}  & \Blue{\widehat{P}=1}  & \Blue{\widehat{P}=2} & 
                \Blue{\widehat{P}=3} & \Blue{\widehat{P}=4} & \Blue{\widehat{P}=5} & 
                \Blue{\widehat{P}=6}\\[.3ex]\hline
\Green{N=3}          &  1    &  3   &  7   & 13  & 22  &  34  &   50\\[.3ex]
\Green{N=4}          &  1    &  4   &  11  & 24  & 46  &  80  &  130\\[.3ex]
\Green{N=5}          &  1    &  5   &  16  & 40  & 86  & 166  &  296\\[.3ex]
\Green{N=6}          &  1    &  6   &  22  & 62  & 148 & 314  &  610\\[.3ex]
\Green{N=7}          &  1    &  7   &  29  & 91  & 239 & 553  & 1163\\[.3ex]\hline
\Red{\rm{components}}    &  1    &  5   &  15  & 35  &  70 & 126  &  210  
\end{array}$$
\caption{Number $n_c(N,\widehat{P})$ of invariant coefficients $T^N_{i_1\ldots i_P}$ for
  $N=3,\ldots,7$ and rank $P\le \widehat{P}=0,\ldots,6$ (rows 2--6) and number $n_t{(\widehat{P})}$ of 
independent tensor components $T^{N,\mu_1\cdots\mu_{P}}$ for rank $P\le \widehat{P}=0,\ldots,6$ (last
  row).}
\label{tab:numbers}
\end{table}

The numbers of invariant coefficients $T^N_{i_1\ldots i_P}$ and tensor
\sloppy
components $T^{N,\mu_1\cdots\mu_{P}}$ 
are compared in \refta{tab:numbers}. For $N\le 4$ the number of invariant coefficients
is smaller than the number of tensor components, which is a basic
precondition of the Passarino--Veltman reduction method. For $N\ge 5$,
on the other hand, the situation is reversed.
Actually, the reduction for $N\ge 6$ presented in (7.7) of
\citere{Denner:2005nn} has been derived in terms of full
tensors. Its translation to tensor coefficients requires a
symmetrization and the resulting coefficients are not unique because
of the redundant number of tensor structures. Therefore, for the
calculation of the tensors $T^{N,\mu_1\cdots\mu_{P}}$ the reduction
for $N\ge 6$ has been implemented in {\collier} also directly at the
tensor level without resorting to a covariant decomposition.

\begin{figure}[t]
  \scalebox{0.95}{
  \begin{picture}(90,70)(-70,0)
    \SetWidth{0.5}
    \SetColor{Black}

    \Text(-70,50)[lb]{\small$B_{i_1\ldots i_P}$}
    \Text(-70,0)[lb]{\small$B^{\mu_1\cdots\mu_P}$}
    \Text(-5,50)[lb]{\small$C_{i_1\ldots i_P}$}
    \Text(-5,0)[lb]{\small$C^{\mu_1\cdots\mu_P}$}
    \Text(60,50)[lb]{\small$D_{i_1\ldots i_P}$}
    \Text(60,0)[lb]{\small$D^{\mu_1\cdots\mu_P}$}
    \Text(125,50)[lb]{\small$E_{i_1\ldots i_P}$}
    \Text(125,0)[lb]{\small$E^{\mu_1\cdots\mu_P}$}
    \Text(190,50)[lb]{\small$F_{i_1\ldots i_P}$}
    \Text(190,0)[lb]{\small$F^{\mu_1\cdots\mu_P}$}
    \Text(255,50)[lb]{\small$G_{i_1\ldots i_P}$}
    \Text(255,0)[lb]{\small$G^{\mu_1\cdots\mu_P}$}
    \Text(320,50)[lb]{\small$\cdots$}
    \Text(320,0)[lb]{\small$\cdots$}    

    \SetWidth{1}
    \LongArrow(-30,55)(-15,55)
    \LongArrow(35,55)(50,55)
    \LongArrow(100,55)(115,55)
    \LongArrow(165,55)(180,55)
    \LongArrow(230,55)(245,55)
    \LongArrow(295,55)(310,55)    

    \SetWidth{1}
    \LongArrow(-55,45)(-55,15)
    \LongArrow(10,45)(10,15)
    \LongArrow(75,45)(75,15)
    \LongArrow(140,45)(140,15)
    \LongArrow(205,45)(205,15)
    \LongArrow(270,45)(270,15)

    \SetWidth{1}
    \LongArrow(165,5)(180,5)
    \LongArrow(230,5)(245,5)
    \LongArrow(295,5)(310,5)   
    
  \end{picture}}
\caption{Reduction chains in {\collier}: For $N\ge 6$ reduction can be
         performed at the tensor level.\label{fig:RedChains}}
\end{figure}
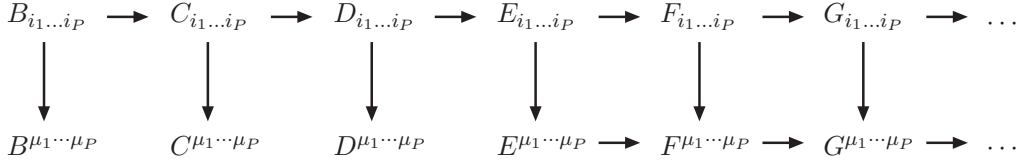

Whereas for $N\le 5$ the recursion exclusively proceeds 
at the coefficient level, 
with $T^{N,\mu_1\cdots\mu_{P}}$ 
constructed afterwards from the respective 
coefficients $T^N_{i_1\ldots i_P}$,
for $N\ge 6$ 
the reduction can alternatively be performed at the level of the tensors.
This means that, in order to calculate a tensor integral with $N\ge 6$,
any $N_{\rm tenred}$ with $5< N_{\rm tenred}\le N$ can be chosen such
that the recursive calculation is performed at the coefficient level for 
$N<N_{\rm tenred}$ and at tensor level for $N\ge N_{\rm tenred}$.
The transition from coefficients to tensors then takes place at $N_{\rm tenred}-1$.
The possible reduction chains are illustrated in \reffi{fig:RedChains}.

\section{Structure of the library}
\label{se:structure}

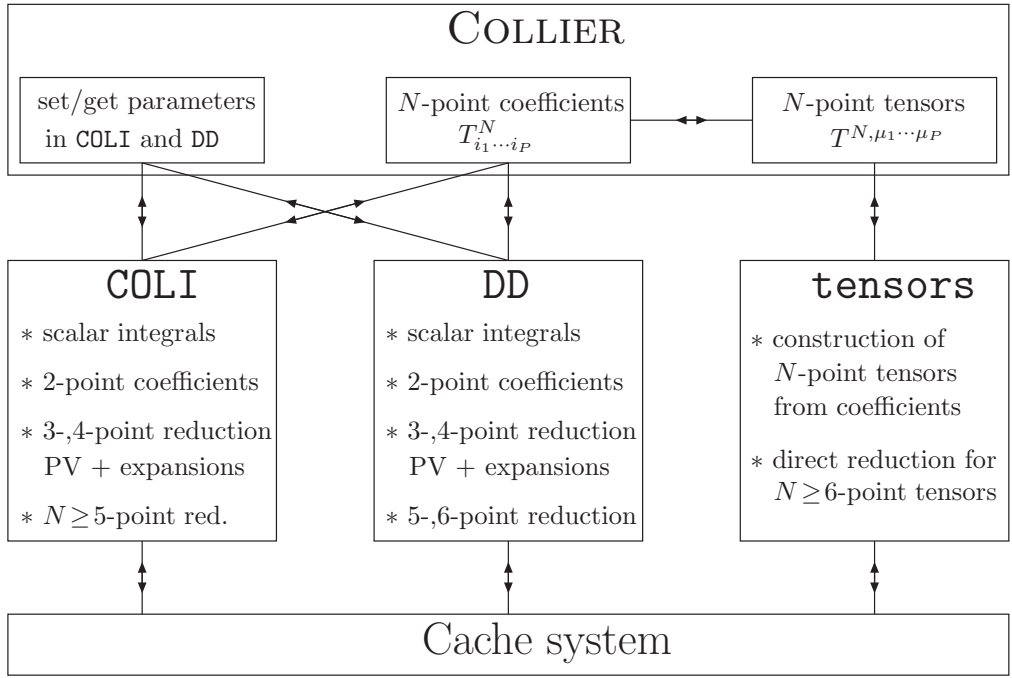
\begin{figure}
  \let\localsize\footnotesize
  \unitlength .92pt\SetScale{0.92} 
  \begin{picture}(200,280)(-80,-30)

    \BBox(-70,180)(340,250)

    \BBox(-65,185)(35,220)
    \BBox(85,185)(185,220)
    \BBox(235,185)(335,220)

    \BBox(-70,30)(40,145)
    \BBox(80,30)(190,145)
    \BBox(230,30)(340,145)
    \BBox(-70,-25)(340,0)


    \LongArrow(-15,30)(-15,10)
    \LongArrow(135,30)(135,10)
    \LongArrow(285,30)(285,10)
    \LongArrow(285,185)(285,160)
    \LongArrow(-15,185)(-15,160)
    \LongArrow(135,185)(135,160)
    \LongArrow(-15,185)(75,161)
    \LongArrow(135,185)(45,161)
    \LongArrow(235,202.5)(205,202.5)

    \LongArrow(-15,0)(-15,20)
    \LongArrow(135,0)(135,20)
    \LongArrow(285,0)(285,20)
    \LongArrow(285,145)(285,170)
    \LongArrow(-15,145)(-15,170)
    \LongArrow(135,145)(135,170)
    \LongArrow(135,145)(45,169)
    \LongArrow(-15,145)(75,169)
    \LongArrow(185,202.5)(215,202.5)

    \Text(110,235)[lb]{\Large{\collier}}

    \Text(-30,130)[lb]{\Large{\coli}}
    \Text(125,130)[lb]{\Large{\DD}}
    \Text(260,130)[lb]{\Large{\tensors}}
    \Text(100,-20)[lb]{\Large{Cache system}}

    \Text(-65,110)[lb]{\localsize{$*$ scalar integrals}}
    \Text(-65,90)[lb]{\localsize{$*$ 2-point coefficients}}
    \Text(-65,70)[lb]{\localsize{$*$ 3-,4-point reduction}}
    \Text(-65,55)[lb]{\localsize{$\;\;$ PV + expansions}}
    \Text(-65,35)[lb]{\localsize{$*$ $N\!\ge\!5$-point red.}}

    \Text(85,110)[lb]{\localsize{$*$ scalar integrals}}
    \Text(85,90)[lb]{\localsize{$*$ 2-point coefficients}}
    \Text(85,70)[lb]{\localsize{$*$ 3-,4-point reduction}}
    \Text(85,55)[lb]{\localsize{$\;\;$ PV + expansions}}
    \Text(85,35)[lb]{\localsize{$*$ 5-,6-point reduction}}

    \Text(235,110)[lb]{\localsize{$*$ construction of}}
    \Text(235,94)[lb]{\localsize{$\;\;$ $N$-point tensors}}
    \Text(235,82)[lb]{\localsize{$\;\;$ from coefficients}}
    \Text(235,60)[lb]{\localsize{$*$ direct reduction for}}
    \Text(235,45)[lb]{\localsize{$\;\;$ $N\!\ge\! 6$-point tensors}}

    \Text(-58,205)[lb]{\localsize{set/get parameters}}
    \Text(-55,192)[lb]{\localsize{in {\coli} and {\DD}}}

    \Text(90,205)[lb]{\localsize{$N$-point coefficients}}
    \Text(115,190)[lb]{\localsize{$T^N_{i_1\cdots i_P}$}}

    \Text(248,205)[lb]{\localsize{$N$-point tensors}}
    \Text(268,192)[lb]{\localsize{$T^{N,\mu_1\cdots \mu_P}$}}
  \end{picture}
\caption{Structure of the library {\collier}.}
\label{fig:structure}
\end{figure}

The structure of {\collier} is illustrated schematically
in \reffi{fig:structure}. The core of the library consists of the
building blocks {\coli} and {\DD}. They constitute two independent
implementations of the scalar integrals $T^N_0$ and the
Lorentz-invariant coefficients $T^N_{i_1\ldots i_P}$ employing the
methods described in the previous section. The building block {\tensors}
provides routines for the construction of the tensors
$T^{N,\mu_1\ldots \mu_P}$ from the coefficients $T^N_{i_1\ldots i_P}$,
as well as for a direct reduction of $N$-point integrals 
for $N\ge 6$ at the tensor
level. The user interacts with the basic routines of {\coli}, {\DD},
and {\tt tensors} via the global interface of {\collier}. It provides
routines to set or extract values of the parameters in
{\coli} and {\DD} as well as routines to calculate the
tensor coefficients $T^N_{i_1\ldots i_P}$ or tensor components
$T^{N,\mu_1\ldots \mu_P}$. The user can choose whether the {\coli} or the
{\DD}~branch shall be used. It is
also possible to calculate each integral with both branches 
and cross-check the results.

In a typical evaluation of a one-loop matrix element, a tensor
integral is called several times with the same kinematical input: 
On the one hand, a single user call
of an $N$-point integral with $P\ge2$ leads to recursive internal
calls of lower $N^{\prime}$-point integrals, 
and for $N^{\prime}\le
N-2$ the same integral is reached through more than one path in the
reduction tree.  On the other hand, different user calls and their
reductions typically involve identical tensor integrals. In order to
avoid multiple calculations of the same integral, the sublibraries of
{\collier} are linked to a global cache system which works as follows:
A parameter $N_{\rm ext}$ numerates external integral calls, while for
the book-keeping of internal calls a binary identifier $id$ is
propagated during the reduction. A pointer is assigned to each index
pair $(N_{\rm ext},id)$. During the evaluation of the first
phase-space points the arguments of the corresponding function calls
are compared, and pairs $(N_{\rm ext},id)$ with identical arguments are
pointed to the same address in the cache. For later phase-space points
the result of the first call of an integral is written to the cache
and read out in subsequent calls pointing to the same address.  Use of
the external cache system is optional and crucially requires that all
calls for tensor integrals are made exactly in the same order for
each phase-space point in a Monte Carlo integration, after an
initialization that signals the beginning of the matrix-element
calculation for the respective phase-space point.
Moreover, internal parameters must not be changed between the first
and the last integral call in each event.

\section{Usage of the library}
\label{subsec:Usage}

\subsection{Installation}
\label{subsec:install}

For the installation of the \collier\ library the package 
{\tt
  collier-$v$.tar.gz} is needed\footnote{The package can be downloaded from 
\href{http://collier.hepforge.org/}{\mbox{http://collier.hepforge.org}}.}
 (where {\tt $v$} stands for the version of the library, \eg {\tt $v=1.0$}),
  and the {\sc CMake} build system
should be installed.  Since \collier\ is a stand-alone Fortran95
code, no additional libraries are required.  

To start installation, {\tt gunzip} and {\tt
  untar collier-$v$.tar.gz} 
  which will unpack into the directory {\tt
  ./COLLIER-$v$} 
  containing the following files and directories
\begin{itemize}
\item {\tt CMakeLists.txt}:  {\sc CMake} makefile to produce the
  \collier\ library,
\item {\tt src:}
\collier\ source directory, containing the main source files  of
\collier\ and further source files in the subdirectories
\begin{itemize}
\item {\tt COLI:} containing the files of the \coli\ branch, 
\item {\tt DDlib:} containing the files of the \DD\ branch,
\item {\tt tensors:} containing the files for tensor construction and
direct tensor reduction,
\item {\tt Aux:} containing auxiliary files, 
\end{itemize}
\item {\tt build:} build directory, where \cmake\ puts all necessary
  files for the creation of the library, such as
object files,
\item {\tt modules:} empty directory for fortran module files,
\item {\tt demos:} directory with demo routines illustrating the use
  of \collier,
\item {\tt COPYING}:  file with copyright information.
\end{itemize}

The \collier\ library is generated by changing 
to the directory {\tt build}
and issuing {\tt "cmake .."} followed by {\tt "make"}:
\cpcsub{%
  {\tt "cd build"}\\*
  {\tt "cmake .."}\\*
  {\tt "make"}\;.
} 
This requires {\sc CMake} to be installed.  Some
information on individual configurations can be found at the top of
the file {\tt CMakeLists.txt}. By default {\tt cmake} sets up the
makefile in such a way
that a dynamic library will be generated. If a static
library is desired, this can be controlled by issuing 
\cpcsub{%
  {\tt "cmake -Dstatic=ON .."} 
} 
in the directory {\tt COLLIER-$v$}.

If no options are specified, {\tt cmake} automatically searches for installed
Fortran compilers and chooses a suited one.
The use of a particular compiler, \eg the {\tt ifort} compiler, can be
enforced by
\cpcsub{%
  {\tt "cmake -DCMAKE\_Fortran\_COMPILER=ifort .."} \;.
} 
The full path to a compiler may be given.

Once the {\tt Makefile} has been generated, the command {\tt make}
will generate the dynamic library {\tt libcollier.so} or the static
library {\tt libcollier.a} in the directory {\tt COLLIER-$v$} which can
be linked to the user's program.

To create the executables for the sample programs (see
\refse{subsec:demo}) in the directory {\tt demos}, the commands
\cpcsub{%
{\tt "make demo"}\\*
{\tt "make democache"}
}
should be issued in the directory {\tt COLLIER-$v$/build}.

All files created by the {\tt "make"} command can be discarded with
\cpcsub{%
  {\tt "make clean"}\\*
  } in the directory {\tt COLLIER-$v$/build}, and all files produced by
{\tt "cmake"} can be eliminated by removing all files in the directory
{\tt COLLIER-$v$/build}.
To clean up completely, all files generated by {\tt "cmake"} should be removed.

\subsection{General usage instructions}
\label{subsec:GenUse}
In order to use \collier\ in a {\tt Fortran} program, the
corresponding modules located in {\tt COLLIER-$v$/modules} have to be
loaded by including the line
\cpcsub{%
  {\tt use COLLIER}\\*
  } 
in the preamble of the respective code, 
and the library
{\tt libcollier.so} or {\tt libcollier.a} in the directory {\tt
  COLLIER-$v$} has to be supplied to the linker.
 This gives access to the
public functions and subroutines of the \collier\ library described in
the following subsections. The names of all these routines end with
the suffix ``{\tt \_cll}''. This name convention is supposed to avoid
conflicts with routine names present in the master program and
increases readability by allowing for an easy identification of
command lines referring to the \collier\ library.

Before \collier\ can be used to calculate tensor integrals, it must be initialized by calling 
\cpcsuptable{character,optional folder\_name xxxx}{%
  \multicolumn{2}{l}{\tt subroutine  Init\_cll(Nmax,rin,folder\_name,noreset)}\\*
  \tt integer Nmax & maximal $\#$ of loop propagators\\*
  \tt integer,optional rin & maximal rank of loop integrals\\*
  \tt character,optional folder\_name & name of folder for output \\*
  \tt logical,optional noreset & no new output folder and files\;,\\
  } 
where the first argument {\tt Nmax} is mandatory, while the other
arguments {\tt rin}, 
{\tt folder\_name} and {\tt noreset} are optional.  With the
argument {\tt Nmax} the user must specify the maximal number $N_{\rm
  max}$ of loop propagators of the tensor integrals $T^{N,P}$ he is going
to calculate ($N\le N_{\rm max}$),
while with the optional second argument {\tt rin} he can specify the maximal
rank $P_{\rm max}$ ($P\le P_{\rm max}$).\footnote{In the {\tt DD} branch of \collier, 
$N$-point integrals are presently only implemented up to $N_{\rm max}=6$, with
5-point functions supported up to rank 5 and 6-point functions supported up to rank 6.
This is sufficient for any $N$-point integral with $N\le 6$ appearing 
in renormalizable theories.}
 If the argument {\tt rin} is omitted, the maximal rank is set to
{\tt Nmax} which is sufficient for integrals in renormalizable theories.
The arguments {\tt Nmax} and {\tt rin} determine the size of internal tables
generated by {\tt Collier}; very large values for these parameters can thus affect the amount 
of allocated memory and the computing time for the integrals. 
As an optional third argument 
the user can pass a string {\tt
folder\_name} to {\tt Init\_cll}, which then creates (or overwrites)
a folder with the specified name in the current directory. The output
of {\collier} will be directed to that folder. If the second
argument is left out, the folder is named {\tt 'output\_cll'} by
default. The creation of an output folder can be suppressed by passing
an empty string {\tt folder\_name=''} to {\tt Init\_cll}. In this case
no output will be created by \collier\ except for the one related to
the initialization and to fatal errors written to the standard output
channel {\tt stdout\_cll=6}. In a second or any subsequent call of
{\tt Init\_cll}, the optional fourth argument {\tt noreset}, 
if present
and set to {\tt .true.}, causes that output folder and files are not
re-created, but that the program continues writing in the existing
files. The optional flag {\tt noreset} is ignored in the very first
call of {\tt Init\_cll}.

The call of {\tt Init\_cll} sets all internal parameters to default values specified in 
\refta{tab:COLLIERparams}. In later calls certain parameters specified in \refse{subsec:ErrOut} are prevented from a reinitialization
if the optional argument {\tt noreset} is flagged {\tt .true.}. After the initialization, many of these parameters can be 
freely set to different values according to the needs of the user. To
this end, \collier\ provides a subroutine {\tt Set}{\it X}{\_cll} for
every parameter {\it X}, and subroutines {\tt SwitchOn}{\it Y}{\_cll},
{\tt SwitchOff}{\it Y}{\_cll} for every flag {\it Y}. To read out the
current value of a parameter {\it X} a routine {\tt get}{\it X}{\_cll}
is available for most parameters. 
The parameters that can be modified by the user and the
respective routines are described in detail in \refse{subsec:SetGet}.
\begin{table}
\begin{center}
\renewcommand{\arraystretch}{1.2}
\renewcommand{\tabcolsep}{.2em}
    \begin{tabular}{c|c|c|c}
parameter  & type   & set with 
           & default  \\
\hline \hline
{\tt mode} & {\tt integer} $\in$ $\{1,2,3\}$ & {\tt SetMode\_cll} 
           & 1 \\
\hline
$\eta_\textrm{req}$ & {\tt double precision} & {\tt SetReqAcc\_cll} 
           & {\tt 1d-8} \\
\hline
$\eta_\textrm{crit}$ & {\tt double precision} & {\tt SetCritAcc\_cll} 
           & {\tt 1d-1} \\
\hline
$\eta_\textrm{check}$ & {\tt double precision} & {\tt SetCheckAcc\_cll} 
           & {\tt 1d-4} \\
\hline
$\mu_\UV^2$ & {\tt double precision} & {\tt SetMuUV2\_cll} 
            & {\tt 1d0} \\
\hline
$\mu_\IR^2$ & {\tt double precision} & {\tt SetMuIR2\_cll} 
            & {\tt 1d0} \\
\hline
$\Delta_\UV$ & {\tt double precision} & {\tt SetDeltaUV\_cll} 
            & {\tt 0d0} \\
\hline
$\Delta_\IR^{(1)},\Delta_\IR^{(2)}$ & {\tt double precision} & {\tt SetDeltaIR\_cll} 
            & {\tt 0d0,0d0} \\
\hline
$\{\overline{m}_{1}^2,\ldots,
\overline{m}_{n_{\textrm{reg}}}^2\}$ & {\tt double complex ($n_\textrm{reg}$)} & {\tt SetMinf2\_cll} 
            & $\{\}$ \\
\hline
$\sigma_\textrm{stop}$ & {\tt integer} $<$ $0$ & {\tt SetErrStop\_cll} 
           & $-8$ \\
\hline
$N_{\rm tenred}$ & {\tt integer} $\ge$ $6$ & {\tt SetTenRed\_cll} 
           & 6 \\
\hline
$n_\textrm{cache}$ & {\tt integer} $\ge$ $0$ & {\tt InitCacheSystem\_cll} 
           & $0$ \\
\hline
$N_{\textrm{cache}}^\textrm{max}$ & {\tt integer ($n_\textrm{cache}$)} $\ge$ $1$ & {\tt SetCacheLevel\_cll} 
           & $-$ \\           
\hline
$n_\textrm{err}$ & {\tt integer} $\ge$ $0$ & {\tt SetMaxErrOut\_cll} 
           & $100$ \\
\hline           
$n_\textrm{err,COLI}$ & {\tt integer} $\ge$ $0$ & {\tt SetMaxErrOutCOLI\_cll} 
           & $100$ \\
\hline
$n_\textrm{err,DD}$ & {\tt integer} $\ge$ $0$ & {\tt SetMaxErrOutDD\_cll} 
           & $100$ \\           
\hline
$n_\textrm{inf}$ & {\tt integer} $\ge$ $0$ & {\tt SetMaxInfOut\_cll} 
           & $1000$ \\           
\hline
$n^{N,\textrm{max}}_\textrm{check}$ & {\tt integer} ($N_\textrm{max}$) $\ge$ $0$ & {\tt SetMaxCheck\_cll} 
           & $\{50,...,50\}$ \\           
\hline
 $n^{B^\prime,\textrm{max}}_\textrm{check}$ & {\tt integer} $\ge$ $0$ & {\tt SetMaxCheckDB\_cll} 
           & $50$ \\           
\hline
$n^{N,\textrm{max}}_\textrm{crit}$ & {\tt integer} ($N_\textrm{max}$) $\ge$ $0$ & {\tt SetMaxCrit\_cll} 
           & $\{50,...,50\}$ \\           
\hline
 $n^{B^\prime,\textrm{max}}_\textrm{crit}$ & {\tt integer} $\ge$ $0$ & {\tt SetMaxCritDB\_cll} 
           & $50$ \\           
\hline
$\widehat{P}^\textrm{max}$ & {\tt integer} $\ge$ $6$ & {\tt SetRitmax\_cll} 
           & $14$ \\           
\hline
{\tt outlev} & {\tt integer}  $\in$ $\{0,1,2\}$  & {\tt SetInfOutLev\_cll} %
           & $2$ \\           
\hline
   \end{tabular}
\end{center}
  \caption{List of \collier\ parameters.}
  \label{tab:COLLIERparams}
\end{table}

After the initialization and a potential redefinition of internal
parameters, \collier\ is ready to calculate tensor integrals. The
generic subroutine {\tt TN\_cll} calculates the coefficients
$T^N_{i_1\dots i_P}$ of the Lorentz-covariant decomposition
\refeq{eq:decomp} of the tensor integrals $T^{N,P}$, while {\tt
  TNten\_cll} returns the 
tensor components $T^{N,\mu_1\dots\mu_P}$.
Alternatively, there are specific subroutines {\tt A\_cll}, {\tt
  B\_cll}, \ldots, {\tt G\_cll} and {\tt Aten\_cll}, {\tt Bten\_cll},
\ldots, {\tt Gten\_cll} for the 1-, 2-, \ldots, 7-point integrals, as well as
{\tt A0\_cll}, {\tt B0\_cll}, \ldots, {\tt D0\_cll}
for the scalar integrals. 
Momentum derivatives of 2-point coefficients,
typically needed to calculate renormalization constants, can be calculated up
to arbitrary rank using the generic subroutine {\tt DB\_cll}, or for the lowest ranks using the
specific subroutines {\tt DB0\_cll}, {\tt DB1\_cll}, {\tt DB00\_cll},
and {\tt DB11\_cll}. More information on the
subroutines for the calculation of tensor integrals is given in \refse{subsec:CalcTen}.

A typical application of \collier\ is to provide 
the one-loop tensor integrals
within an NLO Monte Carlo generator. In this case, the master program performs a loop over 
Monte Carlo events, and for each event it calculates the one-loop matrix element
employing \collier\ to evaluate the corresponding set of tensor integrals. 
In this context, the subroutine
\cpcsuptable{\tt integer,optional cacheNr}{%
\multicolumn{2}{l}{\tt subroutine  InitEvent\_cll(cacheNr)}\\*
\tt integer,optional cacheNr & $\#$ of cache\\ 
}    
should be called for each event before the tensor integrals are computed. 
This call will reinitialize the error flag and the accuracy flag of \collier\ which
can be read out at the end of the sequence of tensor integral calls to obtain global information 
on the status of the calculations. If the cache system is used, the call of {\tt InitEvent\_cll} is
mandatory to reinitialize the cache for every Monte Carlo event. In the case of multiple caches the
respective cache number {\tt cacheNr}
has to be passed to {\tt InitEvent\_cll} as an optional argument. More information
on the use of the cache system can be found in \refse{subsec:cache}.

To help the user to get familiar with the basic usage of \collier,
two sample programs {\tt demo} and {\tt democache} are distributed together with the library. 
They are described in \refse{subsec:demo}.

\subsection{Calculation of tensor integrals}
\label{subsec:CalcTen}

For the tensor integrals, \collier\ provides routines
that deliver the coefficients $T^N_{i_1\dots i_P}$ of the Lorentz-covariant 
decomposition \refeq{eq:decomp} and routines that deliver the 
components of the tensors
$T^{N,\mu_1\ldots\mu_P}$. 

The tensor coefficients $T^N_{i_1\dots i_P}$ are represented by
$N$-dimensional arrays of type {\tt double complex }with the following
convention:
\beq
{\tt TN}(n_0,n_1,n_2,\ldots,n_{N-1})=
 T^N_{\underbrace{\sst 0\ldots0}_{2n_0}
\underbrace{\sst 1\ldots1}_{n_1}
\underbrace{\sst 2\ldots2}_{n_2}
\ldots
\underbrace{\sst N-1\ldots N-1}_{n_{N-1}}}.
\eeq
In this way, all tensor coefficients $T^{N}_{i_1\ldots i_P}$ with
$P=0,\ldots,\widehat{P}$ up to a given rank $\widehat{P}$ are stored
within the same array
\begin{center}
  {\tt double complex TN}($0\!:\![\widehat{P}/2],
                          \underbrace{0\!:\!\widehat{P},\ldots,0\!:\!\widehat{P}}_{N-1}$). 
\end{center}
Note that identical coefficients $T^N_{i_1\ldots i_P}$, related to
each other through a permutation of the indices $\{i_1,\ldots,i_P\}$,
are represented by the same entry of the array {\tt TN}.  As an
example, the mapping between the tensor coefficients $D_{i_1\ldots
  i_P}$ and the
array {\tt D} is explicitly written down in \refta{tab:Dcoefs} 
for the case of the 4-point coefficients up to rank $\widehat{P}=4$. 

Alternatively, the tensor coefficients of the $N$-point integrals up
to rank $\widehat{P}$ can be obtained as a one-dimensional array
\begin{center}
  {\tt double complex TN1}($n_{c}(N,\widehat{P})$), 
\end{center}
where $n_{c}(N,\widehat{P})$ is the total number of coefficients $T^{N}_{i_1\ldots i_P}$
with $i_1\le i_2\le \ldots\le i_P$ and $P\le \widehat{P}$. For $N=1,\ldots,7$ and $\widehat{P}=0,\ldots,6$
the explicit values of $n_{c}(N,\widehat{P})$ are given in \refta{tab:numbers}. The tensor coefficients
are inserted in the array {\tt TN1} with ascending rank $P$ from $P=0$ to $P=\widehat{P}$. 
Coefficients $T^{N}_{i_1\ldots i_P}$ and 
${T}^{N}_{j_1\ldots j_P}$ of equal rank are ordered 
according to the first indices 
$i_k,j_k$ in which they differ. For the 4-point coefficients up to
rank $\widehat{P}=4$, the ordering can be read off from
\refta{tab:Dcoefs}. Note that due to the
{\tt Fortran} limitation of arrays to rank 7, the coefficients of $N$-point integrals with $N\ge 8$ can only be
represented in the format of a one-dimensional array.

\begin{table}
\begin{center}
\renewcommand{\arraystretch}{1.2}
    \begin{tabular}{l|lcl|lcl|l}
$D_0$  & {\tt D}(0,0,0,0)   &\phantom{space}& $D_{113}$ & {\tt D}(0,2,0,1) &\phantom{space}& 
                               $D_{1112}$ & {\tt D}(0,3,1,0) \\  
$D_1$  & {\tt D}(0,1,0,0)   &\phantom{space}& $D_{122}$ & {\tt D}(0,1,2,0) &\phantom{space}& 
                               $D_{1113}$ & {\tt D}(0,3,0,1) \\  
$D_2$  & {\tt D}(0,0,1,0)   &\phantom{space}& $D_{123}$ & {\tt D}(0,1,1,1) &\phantom{space}& 
                               $D_{1122}$ & {\tt D}(0,2,2,0) \\ 
$D_3$  & {\tt D}(0,0,0,1)   &\phantom{space}& $D_{133}$ & {\tt D}(0,1,0,2) &\phantom{space}& 
                               $D_{1123}$ & {\tt D}(0,2,1,1) \\  
$D_{00}$  & {\tt D}(1,0,0,0)   &\phantom{space}& $D_{222}$ & {\tt D}(0,0,3,0) &\phantom{space}& 
                               $D_{1133}$ & {\tt D}(0,2,0,2) \\  
$D_{11}$  & {\tt D}(0,2,0,0)   &\phantom{space}& $D_{223}$ & {\tt D}(0,0,2,1) &\phantom{space}& 
                               $D_{1222}$ & {\tt D}(0,1,3,0) \\  
$D_{12}$  & {\tt D}(0,1,1,0)   &\phantom{space}& $D_{233}$ & {\tt D}(0,0,1,2) &\phantom{space}&
                               $D_{1223}$ & {\tt D}(0,1,2,1) \\ 
$D_{13}$  & {\tt D}(0,1,0,1)   &\phantom{space}& $D_{333}$ & {\tt D}(0,0,0,3) &\phantom{space}&  
                               $D_{1233}$ & {\tt D}(0,1,1,2) \\  
$D_{22}$  & {\tt D}(0,0,2,0)   &\phantom{space}& $D_{0000}$ & {\tt D}(2,0,0,0) &\phantom{space}&
                               $D_{1333}$ & {\tt D}(0,1,0,3) \\  
$D_{23}$  & {\tt D}(0,0,1,1)   &\phantom{space}& $D_{0011}$ & {\tt D}(1,2,0,0) &\phantom{space}&  
                               $D_{2222}$ & {\tt D}(0,0,4,0) \\
$D_{33}$  & {\tt D}(0,0,0,2)   &\phantom{space}& $D_{0012}$ & {\tt D}(1,1,1,0) &\phantom{space}&
                               $D_{2223}$ & {\tt D}(0,0,3,1) \\  
$D_{001}$  & {\tt D}(1,1,0,0)   &\phantom{space}& $D_{0013}$ & {\tt D}(1,1,0,1) &\phantom{space}& 
                               $D_{2233}$ & {\tt D}(0,0,2,2) \\  
$D_{002}$  & {\tt D}(1,0,1,0)   &\phantom{space}& $D_{0022}$ & {\tt D}(1,0,2,0) &\phantom{space}& 
                               $D_{2333}$ & {\tt D}(0,0,1,3) \\  
$D_{003}$  & {\tt D}(1,0,0,1)   &\phantom{space}& $D_{0023}$ & {\tt D}(1,0,1,1) &\phantom{space}& 
                               $D_{3333}$ & {\tt D}(0,0,0,4) \\  
$D_{111}$  & {\tt D}(0,3,0,0)   &\phantom{space}& $D_{0033}$ & {\tt D}(1,0,0,2) &\phantom{space}& 
                               & \\  
$D_{112}$  & {\tt D}(0,2,1,0)   &\phantom{space}& $D_{1111}$ & {\tt D}(0,4,0,0) &\phantom{space}& 
                               &  \\  
   \end{tabular}
\end{center}
  \caption{Mapping between tensor coefficients $D_{i_1\ldots i_P}$ ($P\le 4$) and elements 
${\tt D}(n_0,n_1,n_2,n_3)$ of the array ${\tt D}(0\!:\!2,0\!:\!4,0\!:\!4,0\!:\!4)$. The mapping onto 
the elements of the one-dimensional array representation ${\tt D1}(46)$ is obtained by 
numerating the coefficients in the table starting from the top left entry downwards.   
       }
  \label{tab:Dcoefs}
\end{table}

The full tensor integrals $T^{N,\mu_1\ldots\mu_P}$ are represented by
$4$-dimensional arrays of type {\tt double complex }with the following
convention:
\beq
{\tt TNten}(n_0,n_1,n_2,n_3)=
 T^{N,\overbrace{\sst 0\ldots0}^{n_0}
\overbrace{\sst 1\ldots1}^{n_1}
\overbrace{\sst 2\ldots2}^{n_2}
\overbrace{\sst 3\ldots3}^{n_3}}.
\eeq
In this way, all tensor components $T^{N,\mu_1\ldots \mu_P}$ with
$P=0,\ldots,\widehat{P}$ up to a given rank $\widehat{P}$ are stored
within the same array
\begin{center}
  {\tt double complex TNten}($0\!:\!\widehat{P},0\!:\!\widehat{P},0\!:\!\widehat{P},
                              0\!:\!\widehat{P}$). 
\end{center}
Note that identical components
$T^{N,\mu_1\ldots \mu_P}$, related to
each other through a permutation of the indices $\{\mu_1,\ldots,\mu_P\}$,
are represented by the same entry of the array {\tt TNten}.  The
mapping between the tensor components $T^{\mu_1\ldots \mu_P}$ and the
array {\tt TNten} is explicitly written down in \refta{tab:Telem} up
to rank $\widehat{P}=3$.

Alternatively, the tensor components of the $N$-point integrals up to
rank $\widehat{P}$ can be obtained as a one-dimensional array
\begin{center}
  {\tt double complex TNten1}($n_{t}(\widehat{P})$), 
\end{center}
where $n_{t}(\widehat{P})$ is the total number of tensor components
$T^{N,\mu_1\ldots \mu_P}$
with $\mu_1\le \mu_2\le \ldots\le\mu_P$ and $P\le
\widehat{P}$. For $\widehat{P}=0,\ldots,6$ the explicit values of
$n_{t}(\widehat{P})$ are given in \refta{tab:numbers}. The tensor
components
are inserted in the array {\tt TNten1} with ascending rank $P$ from $P=0$ to $P=\widehat{P}$. 
Components $T^{N,\mu_1\ldots \mu_P}$ and 
${T}^{N,\nu_1\ldots \nu_P}$ of equal rank are ordered 
according to the first indices 
$\mu_k,\nu_k$ in which they differ. For tensors up to rank 
$\widehat{P}=3$, the ordering can be read off from \refta{tab:Telem}.

\begin{table}
\begin{center}
\renewcommand{\arraystretch}{1.2}
\renewcommand{\tabcolsep}{0.4em}
    \begin{tabular}{l|lcl|lcl|l}
$T_0$  & {\tt TNten}(0,0,0,0)   &\phantom{spa}& $T^{22}$ & {\tt TNten}(0,0,2,0) &\phantom{spa}& 
                               $T^{033}$ & {\tt TNten}(1,0,0,2) \\  
$T^0$  & {\tt TNten}(1,0,0,0)   &\phantom{spa}& $T^{23}$ & {\tt TNten}(0,0,1,1) &\phantom{spa}& 
                               $T^{111}$ & {\tt TNten}(0,3,0,0) \\  
$T^1$  & {\tt TNten}(0,1,0,0)   &\phantom{spa}& $T^{33}$ & {\tt TNten}(0,0,0,2) &\phantom{spa}& 
                               $T^{112}$ & {\tt TNten}(0,2,1,0) \\ 
$T^2$  & {\tt TNten}(0,0,1,0)   &\phantom{spa}& $T^{000}$ & {\tt TNten}(3,0,0,0) &\phantom{spa}& 
                               $T^{113}$ & {\tt TNten}(0,2,0,1) \\  
$T^3$  & {\tt TNten}(0,0,0,1)   &\phantom{spa}& $T^{001}$ & {\tt TNten}(2,1,0,0) &\phantom{spa}& 
                               $T^{122}$ & {\tt TNten}(0,1,2,0) \\  
$T^{00}$  & {\tt TNten}(2,0,0,0)   &\phantom{spa}& $T^{002}$ & {\tt TNten}(2,0,1,0) &\phantom{spa}& 
                               $T^{123}$ & {\tt TNten}(0,1,1,1) \\  
$T^{01}$  & {\tt TNten}(1,1,0,0)   &\phantom{spa}& $T^{003}$ & {\tt TNten}(2,0,0,1) &\phantom{spa}&
                               $T^{133}$ & {\tt TNten}(0,1,0,2) \\ 
$T^{02}$  & {\tt TNten}(1,0,1,0)   &\phantom{spa}& $T^{011}$ & {\tt TNten}(1,2,0,0) &\phantom{spa}&  
                               $T^{222}$ & {\tt TNten}(0,0,3,0) \\  
$T^{03}$  & {\tt TNten}(1,0,0,1)   &\phantom{spa}& $T^{012}$ & {\tt TNten}(1,1,1,0) &\phantom{spa}& 
                               $T^{223}$ & {\tt TNten}(0,0,2,1) \\  
$T^{11}$  & {\tt TNten}(0,2,0,0)   &\phantom{spa}& $T^{013}$ & {\tt TNten}(1,1,0,1) &\phantom{spa}& 
                               $T^{233}$ & {\tt TNten}(0,0,1,2) \\  
$T^{12}$  & {\tt TNten}(0,1,1,0)   &\phantom{spa}& $T^{022}$ & {\tt TNten}(1,0,2,0) &\phantom{spa}&
                               $T_{333}$ & {\tt TNten}(0,0,0,3) \\ 
$T^{13}$  & {\tt TNten}(0,1,0,1)   &\phantom{spa}& $T^{023}$ & {\tt TNten}(1,0,1,1) &\phantom{spa}&  
                                   & \\  
   \end{tabular}
\end{center}
  \caption{Mapping between tensor components $T^{\mu_1\ldots \mu_P}$ ($P\le 3$) and elements 
${\tt Tten}(n_0,n_1,n_2,n_3)$ of the array ${\tt TNten}(0\!:\!3,0\!:\!3,0\!:\!3,0\!:\!3)$. The mapping onto 
the elements of the one-dimensional array representation ${\tt TNten1}(35)$ is obtained by 
numerating the coefficients in the table starting from the top left entry downwards.   
       }
  \label{tab:Telem}
\end{table}

The routines for the tensor integrals provide both results for the
full expressions and separately the coefficients of the UV-singular
poles, $1/\eps_\UV$. The latter are useful to determine rational
terms of UV origin. 
Rational terms of IR origin cancel in 
one-loop diagrams~\cite{Bredenstein:2008zb} 
as long as the loop is not inserted into an external line
and only show up in
wave-function renormalization constants which are easily calculated
explicitly.

The coefficients $T^N_{i_1\dots i_P}$ of the Lorentz-covariant
decomposition \refeq{eq:decomp} of the tensor integrals $T^{N,P}$ with
$N=1,\ldots,7$ can be computed by calling the respective subroutines
{\tt A\_cll}, \ldots, {\tt G\_cll}. The argument structure of these
subroutines {\tt {\it N}\_cll}
({\tt {\it N}\_cll}={\tt A\_cll}, \ldots, {\tt G\_cll}) is given by
\cpcsuptable{double complex(${\tt 0\!:\!P0/2},{\tt 0\!:\!P0}, \ldots, 
             {\tt 0\!:\!P0}$) TNuv}{%
\multicolumn{2}{l}{\tt subroutine  {\it N}\_cll(TN,TNuv,MomInv,mass2,R,TNerr)}\\*
\tt double complex(${\tt 0\!:\!R/2},\underbrace{{\tt 0\!:\!R}, \ldots,
  {\tt 0\!:\!R}}_{N-1}$) TN &
    $T^{N,P}_{i_1,\ldots,i_P}$ with $P\le{\tt R}$\\
\tt double complex(${\tt 0\!:\!R/2},\underbrace{{\tt 0\!:\!R},\ldots,{\tt 0\!:\!R}}_{N-1}$) TNuv &
    $T^{N,P\;\UV}_{i_1,\ldots,i_P}$ with $P\le{\tt R}$\\
\tt double complex(1:$n_\mathcal{P}$) MomInv & momentum invariants\\
\tt double complex(0:{\it N-1}) mass2 & squared masses\\
\tt integer R & maximal rank\\ 
\tt double precision(0:R),optional TNerr & error estimates\;.\\
}
The set of $n_\mathcal{P}=\binom{N}{2}$ momentum invariants $\mathcal{P}_N$, represented by the symbolic argument {\tt MomInv}, is ordered such 
that the first $N$ invariants correspond
to the squares $k_i^2$ of the $N$ incoming 
momenta $k_i$, the following $N$ invariants to the squares $(k_i+k_{i+1})^2$ of pairs of adjacent momenta
and so on. In terms of the off-set momenta $p_i=k_1+\ldots+k_i$ entering the 
loop propagators given in \refeq{D0Di}, 
the set of momentum invariants reads
\begin{eqnarray}
  \mathcal{P}_{2k}\!\!\!&=&\!\!\!\left\{ (p_1-p_0)^2,(p_2-p_1)^2,\ldots,(p_{2k-1}-p_{2k-2})^2,
                                     (p_0-p_{2k-1})^2,\right.\nonumber\\
               &&\,(p_2-p_0)^2,(p_3-p_1)^2,\ldots,(p_0-p_{2k-2})^2,(p_1-p_{2k-1})^2,\nonumber\\
               &&\;\;\;\ldots\nonumber\\
               &&\,(p_{k-1}-p_0)^2,(p_k-p_1)^2,\ldots,(p_{k-3}-p_{2k-2})^2,(p_{k-2}-p_{2k-1})^2,\nonumber\\               
               &&\,\left.(p_{k}-p_0)^2,(p_{k+1}-p_1)^2,\ldots,(p_{2k-2}-p_{k-2})^2,(p_{2k-1}-p_{k-1})^2\right\}\!,\hspace{0.8cm}\label{eq:momarg1}\\[1em]
  \mathcal{P}_{2k+1}\!\!\!&=&\!\!\!\left\{ (p_1-p_0)^2,(p_2-p_1)^2,\ldots,(p_{2k}-p_{2k-1})^2,
                                     (p_0-p_{2k})^2,\right.\nonumber\\
               &&\,(p_2-p_0)^2,(p_3-p_1)^2,\ldots,(p_0-p_{2k-1})^2,(p_1-p_{2k})^2,\nonumber\\
               &&\;\;\;\ldots\nonumber\\
               &&\,\left.(p_{k}-p_0)^2,(p_{k+1}-p_1)^2,\ldots,(p_{k-2}-p_{2k-1})^2,
                 (p_{k-1}-p_{2k})^2\right\}\!.\label{eq:momarg2}
\end{eqnarray}
Note that the first $k-1$ lines in \refeq{eq:momarg1} each
contain $N=2k$ elements, while the $k$th line only contains $k=N/2$ elements. In \refeq{eq:momarg2} 
each of the $k$ lines consists of $N=2k+1$ elements. Within each line all momentum indices increase by one unit going from one element to the next one; 
an index that has already reached the maximum value $i=N-1$ passes on to $i=0$. For $N=2,..,7$, the sets of momentum invariants $\mathcal{P}_N$ are explicitly
listed in \ref{App:MomInv}. They have to be provided to the subroutine
{\tt {\it N}\_cll} in terms of parameters of type {\tt double complex}, either as an array of length $n_\mathcal{P}$ or as $n_\mathcal{P}$ single 
parameters. We stress that, although the variable type is {\tt double complex}, non-vanishing imaginary parts of momentum invariants are not yet supported 
by the current version of {\collier}. It is further important to ensure that momentum invariants corresponding to invariant squared masses of single 
external particles assume their exact numerical value, avoiding any deviation that can occur for example if these momentum squares are evaluated numerically. 
Otherwise, problems can appear in IR-divergent integrals where momentum and mass arguments are compared internally in order to decide 
on the correct
analytic expression. Obviously, the argument {\tt MomInv} is absent in the case of one-point integrals {\tt A\_cll}.

The set of squared masses
\begin{equation}
  \mathcal{M}_{N}\,=\,\left\{m_0^2,m_1^2,\ldots,m_{N-1}^2\right\},\label{eq:masses}
\end{equation}
entering the loop propagators given in \refeq{D0Di}, is represented by
$N$ parameters of type {\tt double complex} denoted symbolically as
{\tt mass2}. These parameters with possible non-vanishing (negative)
imaginary parts have to be passed to the subroutine {\tt $N$\_cll}
either as a single array or as individual arguments depending on which
of the two formats (single parameters or array as described above) is
chosen for the momentum invariants {\tt MomInv}.

The {\tt integer} argument {\tt R} represents the maximum rank
$\widehat{P}$ up to which the tensor integrals are calculated.%
It thus defines the size of the output arrays {\tt TN} and {\tt TNuv}
of type {\tt double complex}. As described before, they can be
obtained either as an $N$-dimensional array with the first component
of range $(0\!:\![\widehat{P}/2])$ and the other ones of range
$(0\!:\!{\widehat{P}})$, or as a one-dimensional array of length
$n_c(N,\widehat{P})$.  The respective numbers $n_c(N,\widehat{P})$ are
tabulated by {\tt Collier} during the initialization and can be
obtained with help of the function
\cpcsuptable{\tt integer N,R,nc }{%
\multicolumn{2}{l}{\tt function {GetNc}\_cll(N,R) result(nc)}\\*
\tt integer N,R,nc & $N,\widehat{P},n_c(N,\widehat{P})$\;.\\
}

Finally, there is the option to add an additional output array
{\tt TNerr} consisting of $({0\!:\!{\widehat{P}}})$ entries of type
{\tt double precision} to the list of arguments. If present, the
elements of this array deliver an estimate of the absolute error size
for the tensor coefficients $T^N_{i_1\cdots i_P}$ with all $i_k\neq 0$
of the corresponding rank $P$.  The error estimate $\Delta T^N(P)$ is
determined by following in a rough way the propagation of errors
through the recursion algorithms and by estimating the approximate
size of neglected higher-order terms in the expansions (see
\refse{sec:calc-tensor-coefficients}).  We stress that the returned
values should not be interpreted as precise and reliable error
specifications,
but should rather be regarded as
approximate order-of-magnitude 
estimates on the underlying uncertainties.

Instead of employing the individual subroutines {\tt A\_cll}, \ldots, {\tt
  G\_cll}, tensor coefficients can be calculated for (in principle)
arbitrary $N$ by means of the generic subroutine
\cpcsub{
{\tt subroutine  TN\_cll(TN,TNuv,MomInv,mass2,\nTn,R,TNerr)}\;.\\*
}
The argument structure of the generic {\tt TN\_cll} differs from the
one of the specific {\tt A\_cll}, \ldots, {\tt G\_cll} only by the presence
of the additional {\tt integer} $\nTn$,
\cpcsub{
  {\tt integer \nTn} : \# of loop propagators $(=N)$\;, 
}
defining the number of loop propagators. In the case of the routine
{\tt TN\_cll}, momentum invariants {\tt MomInv}, squared masses {\tt
  mass2}, and resulting coefficients {\tt TN}, {\tt TNuv} can only be
handled in the format of one-dimensional arrays of length
$n_\mathcal{P}$, $N$,
and $n_c(N,\widehat{P})$, respectively.

\begin{sloppypar}
The tensor 
components $T^{N,\mu_1\dots \mu_P}$ of the integrals $T^{N,P}$ with
$N=1,\ldots,7$ can be computed by calling the respective subroutines
{\tt Aten\_cll}, \ldots, {\tt Gten\_cll}. The argument structure of
these subroutines
{\tt {\it N}ten\_cll} = {\tt Aten\_cll}, \ldots, {\tt Gten\_cll} is given by
\cpcsub{
{\tt subroutine}  \\ \qquad {\tt {\it N}ten\_cll(TNten,TNtenuv,MomVec,MomInv,mass2,R,TNtenerr)}\;.\\
}
In addition to the momentum invariants {\tt MomInv} and the squared masses {\tt mass2}, which enter the tensor subroutine {\tt {\it N}ten\_cll} 
exactly in the same way as the coefficient subroutine {\tt {\it N}\_cll}, also the $N-1$ four-vectors $p_i$ 
appearing in the 
loop propagators in \refeq{D0Di} have to be passed to the subroutine {\tt {\it N}ten\_cll}. 
Represented by the symbolic argument {\tt MomVec}, 
\cpcsub{
  {\tt double complex MomVec(\ldots) } : momentum components\;, 
}
they have to be provided either as $N-1$ individual arrays of range
\mbox{$(0\!:\!3)$} corresponding to the momentum components $p_i^\mu$ with
$\mu=0, \ldots, 3$, or combined into a single array of format
$(0\!:\!3,N-1)$.  Note that the same format must be chosen for the
three arguments {\tt MomVec}, {\tt MomInv}, 
and {\tt mass2}, \ie
either all of them must be provided as collections of individual
arguments or all of them must be given as combined arrays. As in the
case of {\tt MomInv}, the variable type is {\tt double complex}
(although non-vanishing imaginary parts are not supported by the
current version of
{\collier}), and the argument {\tt MomVec} is absent for the
one-point integrals {\tt A\_cll}.
\end{sloppypar}

The {\tt integer} argument {\tt R} representing the maximum rank
$\widehat{P}$ up to which the tensor integrals are calculated defines
the size of the output arrays {\tt TNten} and {\tt TNtenuv} of type
{\tt double complex}. As described before, they can be obtained either
as a 4-dimensional array of format
$(0\!:\!\widehat{P},0\!:\!\widehat{P},0\!:\!\widehat{P},0\!:\!\widehat{P})$,
or as a one-dimensional array of length $n_t(\widehat{P})$.
The respective numbers $n_t(\widehat{P})$ are tabulated by {\collier} during the initialization and can be obtained with help of the function
\cpcsuptable{\tt integer N,nt }{%
  \multicolumn{2}{l}{\tt function {GetNt}\_cll(R) result(nt)}\\*
  \tt integer R,nt & $\widehat{P},n_t(\widehat{P})$\;.\\
  }
An error estimate for the tensor components can be accessed
adding the optional output array {\tt TNtenerr} to the list of
arguments. Its entries $(0\!:\!\widehat{P})$ of type {\tt double
  precision} provide estimates on the absolute error size for the
tensor components of the corresponding rank, in a similar manner as in
the case of the subroutine {\tt {\it N}\_cll}.

Instead of employing the individual subroutines {\tt
  Aten\_cll}, \ldots, {\tt Gten\_cll}, tensor components can be calculated
for (in principle) arbitrary $N$ by means of the generic subroutine
\cpcsub{
  {\tt subroutine} \\ \quad {\tt   TNten\_cll(TNten,TNtenuv,MomVec,MomInv,mass2,\nTn,R,TNtenerr)}\;.\\
  } 
The argument structure of the generic {\tt TNten\_cll} differs
from the one of the specific {\tt Aten\_cll}, \ldots, {\tt Gten\_cll}
only by the presence of the additional {\tt integer} $\nTn$ defining
the
number of loop propagators. In the case of the subroutine {\tt
  TNten\_cll}, momenta {\tt MomVec}, momentum invariants {\tt MomInv},
and squared masses {\tt mass2} can only be handled as single arrays of
format $(\mbox{$0\!:\!3$},\mbox{$1\!:\!N-1$})$, $(1\!:\!n_\mathcal{P})$, 
and $(0\!:\!N-1)$, respectively, whereas for the outputs {\tt TNten} and {\tt TNtenuv} 
the user is still free to choose between the array structures 
$(0\!:\!\widehat{P},0\!:\!\widehat{P},0\!:\!\widehat{P},0\!:\!\widehat{P})$ and $(1\!:\!n_t(\widehat{P})$).

Obviously, any call of a coefficient subroutine {\tt A\_cll}, \ldots, {\tt G\_cll} or {\tt TN\_cll}, as well as of a tensor subroutine
{\tt Aten\_cll}, \ldots, {\tt Gten\_cll} or {\tt TNten\_cll}, delivers 
the result for the respective scalar integral within its output array.
A user exclusively interested in the scalar $1$-, \ldots ,$4$-point master
integrals can either restrict the rank argument {\tt R} to
$\widehat{P}=0$, or can
employ the alternative routines {\tt {\it N}0\_cll} = {\tt A0\_cll}, \ldots, {\tt D0\_cll}:
\cpcsub{
{\tt subroutine  {\it N}0\_cll(TN0,MomInv,mass2)}\\
  {\tt double complex TN0 } : $T^N_0$\;.
}
These routines provide the respective scalar integral as a single output variable
{\tt TN0} of type {\tt double complex}, while for the inputs {\tt MomInv} and {\tt mass2} choice can be made
between the usual options. Note that the routines {\tt
  A0\_cll}, \ldots, {\tt D0\_cll} are not linked to the cache system 
and may fail if the Gram determinant for the 3-point function or the
Cayley determinant of the 4-point function vanishes.

Finally, {\tt COLLIER} provides also routines for the calculation of momentum
derivatives of 2-point coefficients, needed for the wave-function
renormalization of external particles. The subroutine in charge, {\tt DB\_cll}, is structured as
\cpcsub{
{\tt subroutine  DB\_cll(DB,DBuv,MomInv,mass2,R,DBerr)}\\
  {\tt double complex DB(\ldots)} \\
  {\tt double complex DBuv(\ldots)} \\
  {\tt double complex DBerr(\ldots)}\;. 
}
The derivatives $B^\prime_{i_1\cdots i_{\widehat{P}}}(p_1^2) \equiv \partial B_{i_1\cdots i_{\widehat{P}}}(p_1^2)/\partial p_1^2$ 
are returned via the output arrays {\tt DB} and {\tt DBuv}, with 
an optional error estimate via {\tt DBerr}. The conventions for the in- and output arguments are in complete
analogy with the ones of the subroutine {\tt B\_cll}. 
The results of the functions $B_0^\prime$, $B_1^\prime$, $B_{00}^\prime$, and $B_{11}^\prime$
can further be obtained as single {\tt double complex} variables with help of the subroutines
\cpcsub{
{\tt subroutine  DB0\_cll(DB0,MomInv,mass2)}\\
  {\tt double complex DB0}\;, \\
{\tt subroutine  DB1\_cll(DB1,MomInv,mass2)}\\
  {\tt double complex DB1}\;, \\
{\tt subroutine  DB00\_cll(DB00,MomInv,mass2)}\\
  {\tt double complex DB00}\;, \\
{\tt subroutine  DB11\_cll(DB11,MomInv,mass2)}\\
  {\tt double complex DB11}\;. \\
}
Since derivatives $B^\prime_{i_1\cdots i_{\widehat{P}}}$ are not cached, 
calls of the subroutines {\tt DB\_cll}, {\tt DB0\_cll},
{\tt DB1\_cll}, {\tt DB00\_cll}, and {\tt DB11\_cll} 
do not interfere with the cache system of {\collier}.

\subsection{Setting and getting parameters}
\label{subsec:SetGet}
The results for the tensor integrals do not only depend on the explicit mass and momentum arguments,
but also on the choices made concerning \textit{regularization parameters}, as well as on
\textit{technical parameters} governing the selection of reduction schemes and the number of iterations for expansion methods.
The parameters of the latter two groups are typically kept at fixed values for a certain set of integral calls. Therefore
they do not form part of the argument lists of the individual integral calls, 
but are rather gathered as a set of global parameters.
They are initialized to default values specified in \refta{tab:COLLIERparams} during the initialization procedure
of {\collier} and can be modified later on. In the following we give details on these parameters, as well as on the subroutines
that allow the user to change or read out their values.

\subsubsection{Regularization parameters}
{\collier} uses dimensional regularization to handle UV divergences. The results for
UV-divergent integrals thus depend on the regulator
$\Delta_{\UV}$ defined in \refeq{eq:normalizationUV} and the scale of dimensional regularization, 
$\mu_\UV$, more precisely on the combination
$\Delta_{\UV}+\ln({\mu^2_\UV}/{Q^2})$, 
where $Q^2$ is some
scale of the respective tensor integral. At fixed order in perturbation theory,
physical S-matrix elements do not depend on $\Delta_{\UV}$ and $\mu_\UV$. In {\collier}, $\Delta_{\UV}$ and $\mu_\UV^2$
are treated as numerical parameters of type {\tt double precision} 
with default values $\Delta_{\UV}=0$ and $\mu_\UV^2=1$, 
which can be modified employing the subroutines
\cpcsub{
{\tt subroutine  SetDeltaUV\_cll(delta)}\\
{\tt double precision delta}\;, \\
{\tt subroutine  SetMuUV2\_cll(mu2)}\\
{\tt double precision mu2}\;. \\
}
On the one hand, by varying these parameters one can verify the 
UV finiteness of S-matrix elements numerically. On the other hand,
in renormalization schemes like MS or $\overline{\textrm{MS}}$, the scale $\mu_\UV$ of dimensional regularization can be identified with the renormalization scale $\mu_\textrm{ren}$
of a running coupling $g(\mu_\textrm{ren})$.
In this case it gains a physical interpretation with impact on S-matrix elements. The current values
of $\Delta_{\UV}$ and $\mu_\UV^2$ can be read out with help of the subroutines
\cpcsub{
{\tt subroutine  GetDeltaUV\_cll(delta)}\\
{\tt double precision delta}\;, \\
{\tt subroutine  GetMuUV2\_cll(mu2)}\\
{\tt double precision mu2}\;. \\
}

By default, also IR divergences are regularized dimensionally. The results of IR-divergent integrals thus depend on $\Delta_{\IR}^{(1)}$ 
and $\Delta_{\IR}^{(2)}$ defined in \refeq{eq:normalizationUV}, and on the scale of dimensional regularization, $\mu_\IR$. 
At fixed order in perturbation theory, IR-finite 
observables do not depend on $\Delta_{\IR}^{(1)}$, $\Delta_{\IR}^{(2)}$, and $\mu_\IR$,
once contributions from virtual and real corrections are combined. In {\collier}, $\Delta_{\IR}^{(1)}$, $\Delta_{\IR}^{(2)}$, and $\mu_\IR^2$
are represented by numerical parameters of type {\tt double precision} 
with default values $\Delta_{\IR}^{(1)}=\Delta_{\IR}^{(2)}=0$ and $\mu_\IR^2=1$,
which can be modified employing the subroutines
\cpcsub{
{\tt subroutine  SetDeltaIR\_cll(delta1,delta2)}\\
{\tt double precision delta1,delta2}\;, \\
{\tt subroutine  SetMuIR2\_cll(mu2)}\\
{\tt double precision mu2}\;. \\
} 
Note, in particular, that $\Delta_{\IR}^{(1)}$ and $\Delta_{\IR}^{(2)}$ 
can be varied independently.  A variation of $\Delta_{\IR}^{(1)}$,
$\Delta_{\IR}^{(2)}$, and $\mu_\IR^2$ can be performed as a
numerical check on the
IR finiteness of observables. 
The current values are retrieved calling
\cpcsub{
{\tt subroutine  GetDeltaIR\_cll(delta1,delta2)}\\
{\tt double precision delta1,delta2}\;, \\
{\tt subroutine  GetMuIR2\_cll(mu2)}\\
{\tt double precision mu2}\;. \\
} 

Collinear divergences can also be regulated introducing a list of mass regulators,
\begin{equation}
  \mathcal{R}_{n_{\textrm{reg}}}\,=\,\left\{\overline{m}_1^{2},\overline{m}_2^{2},\ldots,
  \overline{m}_{n_{\textrm{reg}}}^{2}\right\}.\label{eq:regmasses}
\end{equation}
To this end, the user must call the subroutine
\cpcsub{
{\tt subroutine  SetMinf2\_cll(nminf,minf2)}\\
{\tt double complex minf2(nminf)} \\
{\tt integer nminf}\;, \\
}
where the {\tt integer} variable {\tt nminf} stands for the number
$n_{\textrm{reg}}$ of different regulator masses and the array {\tt
  minf2} contains their squared values $\overline{m}_i^{2}$ of type
{\tt double complex}. Alternatively, regulator masses can be added
successively calling the subroutine
\cpcsub{
{\tt subroutine  AddMinf2\_cll(m2)}\\
{\tt double complex m2}\;, \\
}
which increments $n_{\textrm{reg}}$ by one and adds the {\tt double complex} value {\tt m2} to the list $\mathcal{R}_{n_{\textrm{reg}}}$.
When a tensor integral is called, its arguments (squared masses and momentum invariants) are numerically compared to the elements of 
$\mathcal{R}_{n_{\textrm{reg}}}$. Identified entries are treated as infinitesimally small throughout the calculation 
and their (not necessarily small) numerical values are only kept in otherwise singular logarithms.
It is crucial that the small masses have exactly the same values in the calls of all subroutines.
The number of mass regulators $n_{\textrm{reg}}$ and the list
of their squared values can be read out with
\cpcsub{
{\tt subroutine  GetNminf\_cll(nminf)}\;,\\
{\tt subroutine  GetMinf2\_cll(minf2)}\\
{\tt double complex minf2(nminf)} \\
{\tt integer nminf}\;, \\
}
respectively. Finally, the subroutine
\cpcsub{
{\tt subroutine  ClearMinf2\_cll}\\
}
allows to clear the list $\mathcal{R}_{n_{\textrm{reg}}}$ and to reset $n_{\textrm{reg}}$ to zero.

\subsubsection{Technical parameters}
\label{suse:tecpa}
\collier\ can be run in three different modes that are chosen employing
\cpcsub{
{\tt subroutine  SetMode\_cll(mode)} \\
{\tt integer mode} \\
}
with the {\tt integer} argument {\tt mode=1,2,3}. This subroutine
switches between the different branches implemented.  For {\tt mode=1}
(the default value) the $\tt COLI$ branch is used, for {\tt mode=2}
the $\tt DD$ branch is used, and for {\tt mode=3} the integrals are
calculated in both branches and compared. In the latter case, \collier\ 
returns from the two results the one for which a higher precision is
estimated internally. The error estimate delivered 
(if added as optional argument to the subroutine call)
is determined from the internal estimate of the respective library and
the difference between the $\tt COLI$ and $\tt DD$ results, as the
maximum of these two estimators. Differences between $\tt COLI$ and
$\tt DD$ that exceed a certain threshold value are further reported to
the file {\tt CheckOut.cll} (see \refse{subsec:ErrOut} for details).
The chosen value of {\tt mode} can be retrieved with
\cpcsub{
{\tt subroutine  GetMode\_cll(mode)}\\
{\tt integer mode}\;. \\
}

The target precision $\eta_\textrm{req}$ 
aimed at in the calculation of the tensor integrals can be set calling
\cpcsub{
{\tt subroutine  SetReqAcc\_cll(acc)}\\
{\tt double precision acc} \\
}
with the argument {\tt acc} of type {\tt double precision}. For the
calculation of tensor integrals, {\collier} will choose a reduction
scheme
that is expected to reach the required precision $\eta_\textrm{req}$, if necessary trying 
different choices and performing expansions 
up to the order at which the target precision is achieved.
Hence, the choice of $\eta_\textrm{req}$ affects the precision of the results as well as the run time,
and has to be made in light of the desired balance between the two. 
The default value is $\eta_\textrm{req}=10^{-8}$, and the library has been optimized in particular for this setting. 
The current values of
$\eta_\textrm{req}$ can be inquired with the subroutine
\cpcsub{
{\tt subroutine  GetReqAcc\_cll(acc)} \\
{\tt double precision acc}\;. \\
}
To which extent results
can really be obtained within the precision $\eta_\textrm{req}$ 
depends on the complexity of the problem.
As a second precision threshold, a critical precision
$\eta_\textrm{crit}$, which should be larger than $\eta_\textrm{req}$, 
can be set via the subroutine
\cpcsub{
{\tt subroutine  SetCritAcc\_cll(acc)}\\
{\tt double precision acc}\;. \\
}
The argument {\tt acc} is again of type {\tt double precision}. The critical precision does not influence the actual calculation,
it is a mere book-keeping device: 
If within the sequence of computed integrals for a certain phase-space
point the estimated uncertainty for at least one integral fails to
reach $\eta_\textrm{crit}$, an accuracy flag is
raised to indicate a warning.  The user can consult this flag
at any time and thus dynamically decide how to proceed (\eg if
he wants to discard the respective phase-space point,
 recalculate it with a different branch/different settings, etc.). 
Moreover, critical integrals can be monitored. If this option is selected, their arguments and results are automatically written to an output file. 
More information on the accuracy
flag and the monitoring of critical integrals is given in \refse{subsec:ErrOut}.
The critical precision is initialized as $\eta_\textrm{crit}=10^{-1}$; 
its value can be obtained by  
\cpcsub{
{\tt subroutine  GetCritAcc\_cll(acc)}\\
{\tt double precision acc}\;. \\
}

Finally, a third precision parameter $\eta_\textrm{check}$, which
should be chosen larger than $\eta_\textrm{req}$, governs the
comparison between the results obtained from {\tt COLI} and {\tt DD}.
The default value of this variable is $\eta_\textrm{check}=10^{-4}$;
and it can be modified and read calling 
\cpcsub{
  {\tt subroutine  SetCheckAcc\_cll(acc)}\;,\\
  {\tt subroutine  GetCheckAcc\_cll(acc)} \\
  {\tt double precision acc}\;. \\
  } 
For {\tt mode${=}3$}, integral calls yielding results with a
relative deviation between {\tt COLI} and {\tt DD} of more than
$\eta_\textrm{check}$ are logged in the file {\tt CheckOut.cll}. In
{\tt mode${=}1$} and {\tt mode${=}2$} the parameter $\eta_\textrm{check}$
is irrelevant.

Instead of invoking the described subroutines to set individually the precision thresholds $\eta_\textrm{req}$, $\eta_\textrm{crit}$, and
$\eta_\textrm{check}$, the subroutine 
\cpcsub{
{\tt subroutine  SetAccuracy\_cll(acc0,acc1,acc2)} \\
{\tt double precision acc0,acc1,acc2} \\
}
can be used to set all of them at the same time. 
The {\tt double precision} arguments {\tt acc0}, {\tt acc1}, and {\tt acc2} represent
in this order $\eta_\textrm{req}$, $\eta_\textrm{crit}$, and
$\eta_\textrm{check}$.

A further technical parameter is given by the maximal rank
$\widehat{P}^\textrm{max}$ up to which tensors are calculated in
iterative methods and which thus defines a cut-off order for the
expansion methods. To set $\widehat{P}^\textrm{max}$, the subroutine
\cpcsub{
  {\tt subroutine  SetRitmax\_cll(ritmax)} \\
  {\tt integer ritmax} \\
  } 
can be invoked with the argument {\tt ritmax} bigger or equal to~7.
In turn, the parameter {\tt ritmax} can be retrieved with the help of
\cpcsub{
  {\tt subroutine  GetRitmax\_cll(ritmax)}\;. \\
  } 
The parameter $\widehat{P}^\textrm{max}\ge 7$ acts as maximal rank
for 4-point integrals in the expansion methods, 
the maximal rank for 3- and 2-point integrals
is then internally set to $\widehat{P}^\textrm{max}+2$ and
$\widehat{P}^\textrm{max}+4$, respectively. 
The value of $\widehat{P}^\textrm{max}$ thus affects the precision and
computing time of $N$-point integrals with $N\le 4$ (from external and internal calls), and
the library has been optimized in particular for the  default setting $\widehat{P}^\textrm{max}=14$.
Note further that in order to facilitate the calculation of all tensor integrals $T^{N,P}$ up to
$N= N_{\rm max}$ and $P= P_{\rm max}$ (with $N_{\rm max},P_{\rm max}$ as specified in the initialization
call of {\tt Init\_cll}), $\widehat{P}^\textrm{max}$ cannot be chosen smaller than
$P_{\textrm{max}}+4-N_{\textrm{max}}$.

As explained in \refse{se:methods}, for $N\ge 6$, reduction methods
are implemented in terms of the coefficients $T^{N}_{i_1\dots i_P}$ as
well as in terms the tensor components $T^{N,\mu_1\dots \mu_P}$. A
general calculation of a tensor integral $T^{N,\mu_1\dots \mu_P}$ with
$N\ge 6$ thus proceeds in three steps:
First, for $5\le \bar{N}=N_{\rm tenred}-1\le N$ the coefficients 
$T^{\bar{N}}_{i_1\dots i_{P_{\bar{N}}}}$ 
are calculated recursively
starting from 2-point coefficients. Then 
the tensors $T^{\bar{N},\mu_1\dots \mu_{P_{\bar{N}}}}$ are built from the coefficients 
$T^{\bar{N}}_{i_1\dots i_{P_{\bar{N}}}}$. Finally, the
tensor $T^{N,\mu_1\dots \mu_{P_{N}}}$ is calculated recursively from the tensors 
$T^{\bar{N},\mu_1\dots \mu_{P_{\bar{N}}}}$ (see \reffi{fig:RedChains}
for an illustration).
The threshold $N_{\rm tenred}$ from which on tensor reduction is used can be set and retrieved with the subroutines
\cpcsub{
{\tt subroutine  SetTenRed\_cll(Ntenred)}\;,\\
{\tt subroutine  GetTenRed\_cll(Ntenred)} \\
{\tt integer Ntenred}\;, \\
}
where the argument {\tt Ntenred} 
represents the parameter $N_{\rm tenred}$. The subroutine
\cpcsub{
{\tt subroutine  SwitchOnTenRed\_cll},
}
being equivalent to {\tt SetTenRed\_cll(Ntenred)} with {\tt Ntenred${=}6$},
opts for the maximal level of 
tensor reduction, while the subroutine
\cpcsub{
{\tt subroutine  SwitchOffTenRed\_cll}
}
switches it off completely. The default setting corresponds to maximal tensor reduction $N_{\rm tenred}=6$, 
which is favoured compared to other choices regarding run time.

\subsection{Using the cache system}
\label{subsec:cache} 
\collier\ disposes of a cache system which is used to avoid repeated calculations of identical tensor integrals
and thus serves to speed up computations. It can operate in a local or in a global mode: 
In the local mode,
integrals are only stored during the processing of a single subroutine 
call from \refse{subsec:CalcTen}.
The cache detects identical integrals that 
are reached several times 
in the course of the reduction algorithm 
via different paths
of the reduction tree and avoids their recalculation. In the global mode, in addition also identical integrals belonging 
to different user calls of subroutines from \refse{subsec:CalcTen} are linked. Whereas the local mode of the cache is always at work, 
the global mode has to be activated explicitly. To this end, a number $n_\textrm{cache}$ of separate caches can be created
to store the results of the calculated coefficients or tensors. This is done calling
\cpcsub{
{\tt subroutine  InitCacheSystem\_cll(ncache,Nmax)}\\
{\tt integer ncache,Nmax}\;, \\
}
where the parameters {\tt ncache} and {\tt Nmax} 
represent 
the total number $n_\textrm{cache}$ of caches and the maximal $N=N^\textrm{max}_\textrm{cache}$ up to which
$N$-point integrals are cached, respectively. 
In order to run the cache in the global mode, it is compulsory
that for each phase-space point
the sequence of integral calls assigned to a certain cache {\tt cacheNr} is preceded by a call of {\tt InitEvent(cacheNr)}. 
We further stress that these integral calls must be realized for each 
phase-space point in the same order, and global parameters 
(like $\mu_\UV^2$, the {\tt mode} of \collier, etc.) must not be reset
within a phase-space point, 
because integrals are identified by their position in the sequence of
user calls. Note that there is no hard limit concerning the size of the caches {\tt cacheNr} 
and that the required memory is determined and allocated dynamically during the first phase-space points.
Depending on the application in question,
running the cache
in the global mode can lead to a high use of memory resources.%
\footnote{While a single cache is in principle sufficient, different
  caches are needed if different subprocesses of a particle reaction are calculated
  simultaneously, but not always in exactly the same order.}

Instead of fixing the total number of caches $n_\textrm{cache}$ right from the beginning, it is also possible to subsequently add caches
to the cache system by calling the subroutine
\cpcsub{
{\tt subroutine  AddNewCache\_cll(cache\_no,Nmax)}\\
{\tt integer cache\_no,Nmax}\;. \\
}
The new cache is initialized to store $N$-point integrals up to the value {\tt Nmax} received as input, while the
number assigned to it is returned as output argument {\tt cache\_no}. A call of {\tt AddNewCache\_cll} without a previous initialization
of the cache system is equivalent to the call of  {\tt InitCacheSystem\_cll(ncache,Nmax)} with argument {\tt ncache\,$=1$}.

The threshold $N_\textrm{cache}\le N^\textrm{max}_\textrm{cache}$ up to which integrals are cached can be adjusted individually for each cache. 
For this purpose the subroutine
\cpcsub{
{\tt subroutine  SetCacheLevel\_cll(cache\_no,Nmax)}\\
{\tt integer cache\_no,Nmax} \\
}
is provided.
Note that the level $N^\textrm{max}_\textrm{cache}$ of a cache {\tt cache\_no} can only be changed {\it before} 
the first phase-space point for the respective cache is evaluated
(\ie before {\tt InitEvent\_cll} is evaluated for the first time with the argument
{\tt cache\_no}). Later calls of {\tt SetCacheLevel\_cll} with argument {\tt cache\_no}
are ignored.


With help of the subroutine
\cpcsub{
{\tt subroutine  SwitchOffCacheSystem\_cll}
}
the global cache system can be switched off temporarily. This option
can for instance be useful if an exceptional situation makes it
necessary to depart from the fixed sequence of integrals by inserting
additional integral calls.  The subroutine
\cpcsub{
{\tt subroutine  SwitchOnCacheSystem\_cll}
}
switches the global cache on again,
which will take up its work at the point it was interrupted.

It is also possible to switch off only one particular cache calling
\cpcsub{
{\tt subroutine  SwitchOffCache\_cll(cache\_no)}\\
{\tt integer cache\_no}\;. \\
}
In this case, when switched on again via the subroutine 
\cpcsub{
{\tt subroutine  SwitchOnCache\_cll(cache\_no)}\\
{\tt integer cache\_no}\;, \\
}
the cache either continues at the point where it was paused, or, if in the meantime the subroutine {\tt InitEvent} has
been employed, with the first integral of the list of cache {\tt cache\_no}.

\subsection{Error treatment and output files}
\label{subsec:ErrOut}
Internal errors as well as possible failures in precision are handled by {\collier} in a twofold way: On the one hand,
corresponding global flags for errors and accuracy are set and can be read out by the user, permitting to take measures on the flight.
On the other hand, corresponding error messages and problematic integral calls are recorded in output files.

The error flag $\sigma_\textrm{err}$ is obtained calling
\cpcsub{
{\tt subroutine  GetErrFlag\_cll(errflag)}\\
{\tt integer errflag}\;. \\
}
It is represented by the integer {\tt errflag} assuming values in the
range from $\sigma_\textrm{err}=0$ in the case of faultless processing
to $\sigma_\textrm{err}=-10$ reserved for the most fatal errors.  The
flag $\sigma_\textrm{err}$ keeps its current value until it is
overwritten by a more negative one implying that an error has occurred
that is considered worse than all errors encountered so far. In this
way, $\sigma_\textrm{err}$ indicates the severest error that has
appeared since its initialization. It is automatically reinitialized
to $\sigma_\textrm{err}=0$ with the call of {\tt InitEvent\_cll} for a
new phase-space point, and can in addition be reset at any time
invoking 
\cpcsub{ {\tt subroutine InitErrFlag\_cll}\;.  }  
If $\sigma_\textrm{err}$ falls below a certain threshold
$\sigma_\textrm{stop}$, execution of the program is stopped
automatically.  The default value $\sigma_\textrm{stop}=-8$ is chosen,
so that errors that could be related to the particular characteristics
of an individual phase-space point do not lead to a stop, whereas
systematic errors expected to be common to all phase-space points trigger
the termination of the program. It is possible to select a different
value for $\sigma_\textrm{stop}$ or to retrieve its value, committing 
the corresponding {\tt integer}
argument {\tt stopflag} to the subroutines
\cpcsub{
{\tt subroutine  SetErrStop\_cll(stopflag)}\;,\\
{\tt subroutine  GetErrStop\_cll(stopflag)}\\
{\tt integer stopflag}\;. \\
}
A stop of the program can be suppressed completely upon calling
\cpcsub{
{\tt subroutine  SwitchOffErrStop\_cll()}\;.\\
}

The accuracy flag $\sigma_\textrm{acc}$ works in a very similar manner. 
It serves as an indicator reflecting the precision of the results and
is retrieved as {\tt integer} argument {\tt accflag} via the subroutine
\cpcsub{
{\tt subroutine  GetAccFlag\_cll(accflag)}\\
{\tt integer accflag}\;. \\
}
Initialized with $\sigma_\textrm{acc}=0$, it is changed to
$\sigma_\textrm{acc}=-1$ as soon as an integral calculation does not
reach the target precision $\eta_\textrm{req}$, and to
$\sigma_\textrm{acc}=-2$ if it does not fulfil the critical precision
$\eta_\textrm{crit}$ (see \refse{suse:tecpa} for details on these
parameters). As in the case of the error flag, $\sigma_\textrm{acc}$
is only overwritten by a more negative value and thus signals the
accuracy of the most critical integral calculation since its
initialization. A reinitialization to $\sigma_\textrm{acc}=0$ is
automatically performed with the call of {\tt InitEvent\_cll} for a
new phase-space point, and can in addition be achieved with the help
of the subroutine
\cpcsub{
{\tt subroutine  InitAccFlag\_cll}\;.
}

With the initialization of {\collier}, the user chooses how
messages on errors and accuracy as well as additional information
should be returned. By default, {\collier} stores this information in separate files which are deposited in the directory {\tt output\_cll}
created within the working directory during initialization. 
As described in \refse{subsec:GenUse}, the user can define a different
path or name for the output folder by adding a corresponding string as a second optional argument to the subroutine
{\tt Init\_cll}, or he can suppress the creation of file output by choosing the empty string as folder name. This predefined setting
can be modified later, switching off or on the file output via the subroutines
\cpcsub{
{\tt subroutine  SwitchOffFileOutput\_cll}\;,\\
{\tt subroutine  SwitchOnFileOutput\_cll}\;,
}
or creating a new output directory employing the subroutine
\cpcsub{
{\tt subroutine  SetOutputFolder\_cll(fname)}\\
{\tt character(len=*) fname}\;.\\
}
The path of the output folder is represented by the 
string {\tt fname}, and it can be read out calling
\cpcsub{
{\tt subroutine  GetOutputFolder\_cll(fname)}\\
{\tt character(len=*) fname}\;.\\
}

Error messages are directed to the files {\tt ErrOut.coli}, {\tt
  ErrOut.dd}, and {\tt ErrOut.cll} depending on whether the source of
error is located in the {\tt COLI} library, the {\tt DD} library, or
the global interface (or the module {\tt tensors}), respectively.
During the initialization of {\collier} these files are created
within the above-mentioned output folder,
and a free output channel (with number $>100$) is automatically assigned to
each of them. Output channels can also be attributed manually by the user calling the corresponding subroutines
\cpcsub{
{\tt subroutine  SetnerroutCOLI\_cll(outchan)}\;,\\
{\tt subroutine  SetnerroutDD\_cll(outchan)}\;,\\
{\tt subroutine  Setnerrout\_cll(outchan)}\\
{\tt integer outchan} \\
}
with the channel number {\tt outchan} as {\tt integer} argument. Most notably, this possibility permits to redirect the error output to the standard channel
by choosing {\tt outchan${=}6$}. Note that, if the file output is switched off via the subroutine
{\tt SwitchOffFileOutput\_cll}, the standard channel is not closed and, moreover, {\collier} output directed to it (whether it be the terminal or
a dedicated standard output file) continues to be delivered. The output channels currently selected can be retrieved by the subroutines
\cpcsub{
{\tt subroutine  GetnerroutCOLI\_cll(outchan)}\;,\\
{\tt subroutine  GetnerroutDD\_cll(outchan)}\;,\\
{\tt subroutine  Getnerrout\_cll(outchan)}\\
{\tt integer outchan}\;. \\
}
The number of displayed error messages is limited to $n_{\textrm{err}}^{\textrm{max}}=100$ by default in order to protect the error files from
growing in size without control. This predefined limit can be modified individually for the three types of errors with help of the subroutines
\cpcsub{
{\tt subroutine  SetMaxErrOutCOLI\_cll(nout)}\;,\\
{\tt subroutine  SetMaxErrOutDD\_cll(nout)}\;,\\
{\tt subroutine  SetMaxErrOut\_cll(nout)}\\
{\tt integer nout} \\
}
specifying a corresponding {\tt integer} number {\tt nout}. By default
the error limits as well as the respective counters 
are reset in case of a reinitialization of {\collier}, but this reset
is suppressed if {\tt noreset$=$.true.}\ is passed as 
additional argument to {\tt Init\_cll}. The counters can be
reinitialized by hand calling 
\cpcsub{
{\tt subroutine  InitErrCntCOLI\_cll}\;,\\
{\tt subroutine  InitErrCntDD\_cll}\;,\\
{\tt subroutine  InitErrCnt\_cll}\;.
}
The error output can be dis- and enabled by calling
\cpcsub{
{\tt subroutine  SetErrOutLev\_cll(outlev)}\\
{\tt integer outlev} \\
}
with {\tt integer} argument {\tt outlev${=}0$} and {\tt outlev${=}1$}, respectively. In case {\collier} is initialized with an empty string 
as output-folder name, error output is disabled by default, while in all other cases it is switched on.

Additional information and status messages not related to errors are
recorded in the log-file 
{\tt InfOut.cll}, created as well during the 
initialization of {\collier} in the designated output folder. Also in this case a free output channel (with number $>100$) is automatically 
generated, and can be adapted by the user passing the corresponding {\tt integer} number {\tt outchan} to the subroutine
\cpcsub{
{\tt subroutine  Setninfout\_cll(outchan)}\\
{\tt integer outchan}\;. \\
}
Again, this allows for a redirection of the output to the standard channel {\tt outchan${=}6$}. 
To retrieve the current output channel, the subroutine
\cpcsub{
{\tt subroutine  Getninfout\_cll(outchan)}\\
{\tt integer outchan} \\
}
can be called. By default the output is limited to $n_{\textrm{inf}}^{\textrm{max}}=1000$ messages,
but the user can modify this predefined limit employing
\cpcsub{
{\tt subroutine  SetMaxInfOut\_cll(nout)}\\
{\tt integer nout}\;. \\
} 
By default $n_{\textrm{inf}}^{\textrm{max}}$
as well as the respective counter
are reset in case of a reinitialization of {\collier}, unless this reset
is suppressed by passing  {\tt noreset$=$.true.}\ as 
additional argument to {\tt Init\_cll}.
To which extent informative output is provided, can be regulated by a
call of the subroutine
\cpcsub{
{\tt subroutine  SetInfOutLev\_cll(outlev)} \\
{\tt integer outlev} \\
}
with a suitable {\tt integer} argument {\tt outlev${=}0,1,2$}.
Initialization of {\collier} with an empty string as folder name,
implies a presetting of {\tt outlev${=}0$} (disabled output), while in
all other cases the presetting {\tt outlev${=}2$} (maximum output) is
selected. In the latter case, any change of an internal parameter is
recorded in the output file which can lead to excessive output for
instance if some parameters (such as the UV scale $\mu_\UV^2$) are
modified repeatedly (\eg for each phase-space point). Hence, in
addition an intermediate output level {\tt outlev${=}1$} is offered,
tracing only more special activities which are considered to occur
less frequently.

As soon as the mode in which {\collier} operates is switched for the
first time to {\tt mode${=}3$}, the file {\tt CheckOut.cll} is created
in the common output folder. The channel automatically assigned to the
output file can be changed to a different number, and the current
channel number can be retrieved, calling the subroutines
\cpcsub{
{\tt subroutine  Setncheckout\_cll(outchan)}\;,\\
{\tt subroutine  Getncheckout\_cll(outchan)}\\
{\tt integer outchan}\;, \\
}
respectively. In {\tt mode${=}3$}, integrals are calculated both with
the {\tt COLI} and the {\tt DD} branch of the library, and the file
{\tt CheckOut.cll} collects input and results for those integrals with
a relative discrepancy of more than $\eta_\textrm{check}$ (see
\refse{suse:tecpa} for more details on the parameter
$\eta_\textrm{check}$). Also derivatives of 2-point functions are
compared between {\tt COLI} and {\tt DD} and reported in case of
relative deviations above $\eta_\textrm{check}$.  Output is limited to
the first $n^{N,\textrm{max}}_\textrm{check}$ problematic $N$-point
integrals and the first $n^{B^\prime,\textrm{max}}_\textrm{check}$
derivatives.  The limit $n^{N,\textrm{max}}_\textrm{check}$ can be
individually chosen for each $N$ via the subroutine
\cpcsub{
{\tt subroutine  SetMaxCheck\_cll(npoints,N)}\\
{\tt integer npoints,N}\;, \\
}
where the 
inputs {\tt npoints} and {\tt N} represent $n^{N,\textrm{max}}_\textrm{check}$ and $N$, respectively. The limit
$n^{B^\prime,\textrm{max}}_\textrm{check}$ for the derivatives of 2-point functions can be chosen in an analogous manner employing the subroutine
\cpcsub{
{\tt subroutine  SetMaxCheckDB\_cll(npoints)}\\
{\tt integer npoints}\;. \\
}
It is also possible to set all $n^{1,\textrm{max}}_\textrm{check}$, \ldots, $n^{N_{\textrm{max}},\textrm{max}}_\textrm{check}$ 
with a single subroutine call,
transmitting the {\tt integer} array $\{n^{1,\textrm{max}}_\textrm{check},\ldots,n^{N_{\textrm{max}},\textrm{max}}_\textrm{check}\}$
as a single argument {\tt npointarray} to
\cpcsub{
{\tt subroutine  SetMaxCheck\_cll(npointarray)}\\
{\tt integer npointarray ($N_\textrm{max}$)}\;. \\
}
Note that the value $N_{\textrm{max}}$ is fixed from the call of {\tt
  Init\_cll} of the last \mbox{(re-)}initialization of {\collier} (see
\refse{subsec:GenUse}).  Initial values are
$n^{1,\textrm{max}}_\textrm{check}=\ldots=n^{N_{\textrm{max}},\textrm{max}}_\textrm{check}=n^{B^\prime,\textrm{max}}_\textrm{check}=50$,
and the call of {\tt Init\_cll} leads to a reinitialization of the
limits and the respective counters unless the optional argument {\tt
  noreset} is present and {\tt .true.}. We further remark that the
counters for the output messages in {\tt Checkout.cll} can also be
reset to zero by hand calling
\cpcsub{
{\tt subroutine  InitCheckCnt\_cll}
}
for the $N$-point integrals, and 
\cpcsub{
{\tt subroutine  InitCheckCntDB\_cll}
}
for the 2-point derivatives.

  Finally, the user can request additional information about integrals
  for which the estimated precision does not reach the threshold
  $\eta_\textrm{crit}$ (see \refse{suse:tecpa} for more details on the
  parameter $\eta_\textrm{crit}$). This feature has to be launched
  explicitly via 
\cpcsub{ {\tt call InitMonitoring\_cll}\;.  }  
After activation, input and results for integrals that fall short in
precision are recorded to the file {\tt CritPointsOut.cll}. As for the
other files, a free output channel is automatically attributed.  The
channel number can be set by hand and read out with the help of the
subroutines
\cpcsub{
{\tt subroutine  Setncritpointsout\_cll(outchan)}\;,\\
{\tt subroutine  Getncritpointsout\_cll(outchan)}\\
{\tt integer outchan}\;. \\
} 
Output is limited to the first $n^{N,\textrm{max}}_\textrm{crit}=50$
problematic $N$-point integrals and the first
$n^{B^\prime,\textrm{max}}_\textrm{crit}=50$ derivatives.  The limits
$n^{N,\textrm{max}}_\textrm{crit}$ and
$n^{B^\prime,\textrm{max}}_\textrm{crit}$ can be changed making
use of the subroutines
\cpcsub{
{\tt subroutine  SetMaxCritPoints\_cll(npoints,N)}\;,\\
{\tt subroutine  SetMaxCritPoints\_cll(npointarray)}\;,\\
{\tt subroutine  SetMaxCritPointsDB\_cll(npoints)}\\
{\tt integer npoints,N} \\
{\tt integer npointarray($N_\textrm{max}$)}\;, \\
}
which work in a completely analogous manner as the subroutines
{\tt SetMaxCheck\_cll} and {\tt SetMaxCheckDB\_cll} described above.
Initialization or reinitialization of {\collier} does not change 
$n^{N,\textrm{max}}_\textrm{crit}$,
$n^{B^\prime,\textrm{max}}_\textrm{crit}$ and the respective counters.
However, a call of {\tt InitMonitoring\_cll}
resets
$n^{1,\textrm{max}}_\textrm{crit}=\ldots=n^{N_{\textrm{max}},\textrm{max}}_\textrm{crit}=n^{B^\prime,\textrm{max}}_\textrm{crit}=50$,
and the respective counters to zero. 
A reset of only the counters can be enforced by hand calling
\cpcsub{
{\tt subroutine  InitPointsCnt\_cll}\;.\\
}

\subsection{Sample programs}
\label{subsec:demo}
With the commands
\cpcsub{%
{\tt "make demo"}\\*
{\tt "make democache"}\\*
}
in the directory {\tt COLLIER-$v$/build}
two sample programs can be installed that can be run executing
\cpcsub{%
{\tt "./demo"}\\*
{\tt "./democache"}\\*
}
in the folder {\tt COLLIER-$v$/demos}.

The program {\tt demo} is dedicated to the calculation of single
tensor integrals. During the run, the user is asked to specify the
mode and to choose among various examples $X$ of $N$-point integrals
the one he likes to compute. The result of the calculation is written
to a file {\tt demo\_$N$point\_example$X\/$.dat}, which directs the
user to the passage within {\tt demo.f90} where the source code for
the respective integral call can be found. In many cases, various
calls of the same integral are shown featuring the different options
for communicating the arguments to the subroutine.  The variables used
in the examples are defined at the beginning of the file {\tt
  demo.f90}, followed by a short sequence of code common to all
examples where {\collier} is initialized. It contains various
commented lines which can be activated removing the exclamation mark
in front and which demonstrate how global parameters of {\collier}
can be modified.

The program {\tt democache} exemplifies the usage of the cache. For a
set of 1000 phase-space points a series of 8 tensor integral
computations is performed several times. The toy Monte Carlo is
carried out subsequently in four different subsets: first using the
{\tt COLI} branch, with and without cache, then using the {\tt DD}
branch, with and without cache. The source code is located in the file
{\tt democache.f90}.

\section{Conclusions}
\label{se:concl}

  The fortran-based library \collier\ numerically evaluates
  one-loop scalar and tensor integrals in perturbative relativistic
  quantum field theories for scattering processes with no a-priori
  restriction on the particle multiplicities.
  The particular strengths of \collier\ comprise the use of 
  dedicated techniques to automatically optimize numerical stability 
  in delicate phase-space regions, the support of complex internal masses
  for unstable particles,
  and the optional use of dimensional or mass regularization 
  to treat infrared divergences.
  Moreover, \collier\ allows for powerful checks on the correctness
  and numerical stability of the results, since it is a merger
  of two independent integral libraries, $\tt COLI$ and $\tt DD$.

  \collier\ can be used both within traditional Feynman-diagrammatic
  and modern unitarity-based calculations, delivering all relevant
  scalar and tensor one-loop integrals on demand. The library 
  already represents an essential building block in the automated
  one-loop amplitude generators {\sc OpenLoops} and {\sc Recola}
  and is now ready to be employed by other generators as well.

\section{Acknowledgements}
We thank B.~Biedermann, F.~Cascioli, R.~Feger, J.N.~Lang, J.~Lindert,
P.~Maierh\"ofer, M.~Pellen, S.~Pozzorini, A.~Scharf and S.~Uccirati
for performing various checks of the code.  We are further grateful to
J.N.~Lang for providing the {\sc CMake} makefile for the library.
This work was supported in part by the Deutsche Forschungsgemeinschaft
(DFG) under reference number DE~623/2-1.  The work of L.H.\ was
supported by the grants FPA2013-46570-C2-1-P and 2014-SGR-104, and
partially by the Spanish MINECO under the project MDM-2014-0369 of
ICCUB (Unidad de Excelencia ``Mar\'ia de Maeztu'').

\appendix
\section{Sets of momentum invariants $\mathcal{P}_N$ for $N=1,\ldots,7$}
\label{App:MomInv}
The $N$-point tensor integrals depend on the complete set of momentum
invariants $\mathcal{P}_N$ that can be formed
from the momenta $p_i$ entering the propagator denominators in
\refeq{D0Di}. Our convention for the order of elements within the sets
$\mathcal{P}_N$ is
given in \refeq{eq:momarg1} and \refeq{eq:momarg2}. For convenience, we list here
explicitly the $\mathcal{P}_N$ for $N=2,\ldots,7$:
\begin{eqnarray}
  \mathcal{P}_{2}\!\!\!&=&\!\!\!\left\{p_1^2\right\}\!,\nonumber\\[1.5ex]
  \mathcal{P}_{3}\!\!\!&=&\!\!\!\left\{p_1^2,\,(p_2-p_1)^2,\,p_2^2\right\}\!,\nonumber\\[1.5ex]
  \mathcal{P}_{4}\!\!\!&=&\!\!\!\left\{p_1^2,\,(p_2-p_1)^2,\,(p_3-p_2)^2,\,p_3^2,\,p_2^2,\,(p_3-p_1)^2\right\}\!,\nonumber\\[1.5ex]
  \mathcal{P}_{5}\!\!\!&=&\!\!\!\left\{p_1^2,\,(p_2-p_1)^2,\,(p_3-p_2)^2,\,(p_4-p_3)^2,\,p_4^2,\right.\nonumber\\
                       && \left.p_2^2,\,(p_3-p_1)^2,\,(p_4-p_2)^2,\,p_3^2,\,(p_1-p_4)^2\right\}\!,\nonumber\\[1.5ex]
  \mathcal{P}_{6}\!\!\!&=&\!\!\!\left\{p_1^2,\,(p_2-p_1)^2,\,(p_3-p_2)^2,\,(p_4-p_3)^2,\,(p_5-p_4)^2,\,p_5^2,\right.\nonumber\\
                       && \left.p_2^2,\,(p_3-p_1)^2,\,(p_4-p_2)^2,\,(p_5-p_3)^2,\,p_4^2,\,(p_1-p_5)^2,\right.\nonumber\\
                       && \left.p_3^2,\,(p_4-p_1)^2,\,(p_5-p_2)^2\right\}\!,\nonumber\\[1.5ex]
  \mathcal{P}_{7}\!\!\!&=&\!\!\!\left\{p_1^2,\,(p_2-p_1)^2,\,(p_3-p_2)^2,\,(p_4-p_3)^2,\,(p_5-p_4)^2,\,(p_6-p_5)^2,\,p_6^2,\right.\nonumber\\
                       && \left.p_2^2,\,(p_3-p_1)^2,\,(p_4-p_2)^2,\,(p_5-p_3)^2,\,(p_6-p_4)^2,\,p_5^2,\,(p_1-p_6)^2,\right.\nonumber\\
                       && \left.p_3^2,\,(p_4-p_1)^2,\,(p_5-p_2)^2,\,(p_6-p_3)^2,\,p_4^2,\,(p_1-p_5)^2,\,(p_2-p_6)^2\right\}\!.\hspace{1cm}
\end{eqnarray}



\bibliographystyle{elsarticle-num}

\bibliography{collier}







\end{document}